\documentclass[10pt,twocolumn,twoside]{IEEEtran}

\usepackage{array}
\newcolumntype{P}[1]{>{\centering\arraybackslash}p{#1}}
\usepackage{float}
\usepackage{tabu}
\usepackage{amsmath}
\usepackage{graphicx}
\usepackage{blindtext}
\usepackage{epsfig} 
\usepackage{mathptmx} 
\usepackage{times} 
\usepackage{amssymb}  
\usepackage{multirow}
\usepackage{subcaption}
\usepackage{csquotes}
\usepackage{algpseudocode}
\usepackage{algorithm}
\usepackage{tikz}
\usetikzlibrary{shapes.geometric, arrows}
\usepackage{longtable}
\usepackage{amsbsy}

\usepackage{authblk}
\usepackage{leftidx}
\usepackage{ragged2e}
\usepackage[font={scriptsize}]{caption}
\usepackage{color}

\begin{document}

\title{Attack Analysis and Resilient Control Design for Discrete-Time Distributed Multi-Agent Systems}


\author{Aquib Mustafa, \textit{Student Member, IEEE} and Hamidreza Modares, \textit{Senior Member, IEEE} 
\thanks{
Aquib Mustafa and Hamidreza Modares are with  the Department
of Mechanical Engineering, Michigan State University, East Lansing, MI, 48863, USA (e-mails:mustaf15@msu.edu; modaresh@msu.edu).}
}

\maketitle

 \thispagestyle{empty}
 \pagestyle{empty}

\begin{abstract}
This work presents a rigorous analysis of the adverse effects of cyber-physical attacks on discrete-time distributed multi-agent systems, and propose a mitigation approach for attacks on sensors and actuators. First, we show how an attack on a compromised agent can propagate and affect intact agents that are reachable from it. That is, an attack on a single node snowballs into a network-wide attack and can even destabilize the entire system. Moreover, we show that the attacker can bypass the robust $H_{\infty}$ control protocol and make it entirely ineffective in attenuating the effect of the adversarial input on the system performance. Finally, to overcome adversarial effects of attacks on sensors and actuators, a distributed adaptive attack compensator is designed by estimating the normal expected behavior of agents. The adaptive attack compensator is augmented with the controller and it is shown that the proposed controller achieves secure  consensus in presence of the attacks on sensors and actuators. This controller does not require to make any restrictive assumption on the number of agents or agent's neighbors under direct effect of adversarial input. Moreover, it recovers compromised agents under actuator attacks and avoids propagation of attacks on sensors without removing compromised agents. The effectiveness of the proposed controller and analysis is validated on a network of Sentry autonomous underwater vehicles subject to attacks under different scenarios.
\end{abstract}


\IEEEpeerreviewmaketitle

\section{Introduction}

A cyber-physical system (CPS) refers to a class of engineering systems that integrates the cyber aspect of computation and communication elements with physical entities. Based on their control objectives CPSs can be categorized into two classes, namely distributed multi-agent systems (DMASs) and centralized networked control systems (CNCSs). The control objective in DMAS is to achieve a coordinated or synchronized motion or behavior through the exchange of local information among agents \cite{c6}-\cite{c8}. On the other hand, the control objective in CNCS, for which the feedback loops are closed through a communication network, is to regulate the system's output to a desired value or trajectory \cite{c2}-\cite{c5}. DMASs and CNCSs are both prone to cyber-physical attacks and corruption of sensory data or manipulation of actuators' inputs which can severely and adversely affect their performance.\par
Stealthy attacks in CNCSs are considered as attacks that significantly disrupt the systems' states while assuring that the system outputs, observed by the system monitors, remain within their acceptable bounds. On the other hand, the bulk of the work on the resilient control of DMASs assumes that agents exchange their states, and not outputs, with each other over a communication network to achieve consensus or synchronization. In this class of systems, the effects of adversaries are analyzed based on the discrepancy between the state of agents and their neighbors. A stealthy attack on the communication network can remain unnoticed, but attacks on sensors and actuators can be detected if the agent's states do not follow its system dynamics. However, mitigation of attacks without removing them and harming the network connectivity is not straightforward, and, as shown in this paper, if the mitigation process does not take actions fast, the entire network can become unstable.\par
Considerable results have been presented for detection  \cite{c13}-\cite{c271} and mitigation of attacks in DMAS. There are generally two approaches in designing mitigation techniques for DMAS. In the first approach, a monitor is designed to detect attacks on neighbors and then remove compromised agents, once identified \cite {c32}-\cite {c36}. In these approaches, each normal agent either uses an observer for each of its neighbors to detect abnormality  \cite{c16} or discard neighbors information based on the discrepancy between actual and malicious agents using an iterative strategy \cite{c32}-\cite{c33}. The former approach requires a model for each of its neighbors which makes it not scalable. The latter requires meeting the $F$-total or the $F$-local condition.  That is, there should be an upper bound on $F$ either on the total number of adversarial agents, called as $F$-total or on the local number of compromised agents in the neighborhood of each intact agent, called as $F$-local. Although these approaches can counteract variety of attacks, including attacks on sensors, actuators and communication network, they might harm the network connectivity by rejecting neighbor's information even if there is no attack. This is because they might not be able to distinguish between a change in neighbors behavior due to attack and a legitimate change in the system. For example, in a leader-follower synchronization problem, a legitimate change in leader's state can be detected as a change due to adversarial input by neighbors. Moreover, these approaches treat all types of attacks the same by discarding compromised agents. However, as shown in this paper, attacks on sensors and actuators can be recovered and compromised agents can be brought back to the network without making any restrictive assumption on the network connectivity. This avoids any unnecessary harm to the network connectivity.\par  

In the second approach, local resilient control protocols based attack mitigation are designed to directly mitigate attack without identifying them. Reputation-based resilient control protocol is presented in \cite {c43} for leader-follower problem under certain conditions. Game-theory based resilient control architectures \cite {c401}-\cite {c404} are presented to minimize the effects of adversarial input. With an assumption of having partial knowledge of attacker, a resilient receding horizon-based control protocol is discussed in \cite {c38}-\cite {c39} for mitigation of the replay attack. Secure state estimation and control under sensor attack is considered in \cite {c30}-\cite {c301}. A resilient control protocol is presented in \cite {c42} for single and double integrator system based on local state emulator. In \cite {c371}, an adaptive resilient control protocol is presented for the attack on sensor and actuator of the system. Most of these results are presented for continuous-time systems. However, in real-time applications,  the system communicates and broadcasts there information at discrete instants.\par

To design a resilient control protocol, one needs to identify the adverse effects of the attack on the system performance from the attacker's perspective. Despite tremendous progress in identifying the adverse effects of attacks on DMAS, there is still a need to identify the vast effects of stealthy attacks and equipt the system with resilient control protocol to mitigate them. Toward this end, in this paper,  first, we illustrate how an attack on a compromised agent spreads across the network and affects intact agents that are reachable from a compromised agent. Then, we show that the attacker can design a stealthy attack that has a common mode with the system dynamics and launch on a single root node to destabilize the entire system. We call this as the internal model principle for the attacker in discrete-time DMAS.  The attacker does not need to know the graph topology or agents dynamics to design its attack signal and can eavesdrop on some sensory information to identify one eigenvalue of the consensus dynamics. We also show that the attacker can entirely disable robust techniques such as $H_{\infty}$, used for attenuating the effects of adversarial inputs on the performance of the system.

To mitigate the effect of the adversarial input, this work presents a distributed adaptive resilient controller. First, the expected normal behavior of agents are predicted using an observer-like dynamics. Then, an adaptive attack compensator is designed using predicted normal behavior of agents. The designed adaptive attack compensator is augmented with the controller for the mitigation of the attack.  Moreover, we have shown that the consensus error is uniformly bounded using the proposed controller in the presence of the attack. The proposed adaptive resilient control protocol makes no restriction on graph topology as compared to the existing approaches \cite{c32}-\cite{c36}. The proposed controller preserves the network connectivity and mitigates the effect of adversarial input on the actuator of the compromised agent. That is, not only the synchronization is achieved in the presence of actuator attacks, but also compromised agents are recovered. On other the hand, attacks on sensor affect only compromised agents without being propagated in the network. Finally, simulation results validate the effectiveness of the proposed controller and theoretical analysis for a network of Sentry autonomous underwater vehicles under the influence of attacks for different scenarios.

\vspace{-0.3cm}
\section{Notations and Preliminaries}
In this section, the preliminaries of graph theory and standard distributed consensus of multi-agent systems are provided. 

\vspace{-0.2cm}

\subsection{Graph Theory}
A directed graph $\mathcal{G}$ consists of a pair $(\mathcal{V,\,{\mathcal{E}}})$ in which set of nodes and set of edges are represented by $\mathcal{V}={v_1,\dots,v_N}$ and  ${\mathcal{E}} \subset \mathcal{V}$x$\mathcal{V}$, respectively. The adjacency matrix is defined as $\mathcal{A}=[{a}_{ij}]$, with $a_{ij}>0$ if $(v_j, v_i) \in \mathcal{E}$. The set of  nodes $v_i$ with edges incoming to node $v_j$ is called as neighbors of node $v_i$, namely $\mathcal{N}_i={v_j:(v_j,v_i) \in \mathcal{E}}$. The graph Laplacian matrix is defined as $L=H-\mathcal{A}$, where $H=diag(h_i)$ is known as the in-degree matrix, with $\sum\nolimits_{j \in N_i} a_{ij}$ as the weighted in-degree of node $i$. A node is called as a \textit{root node} if it can reach all other nodes of the graph $\mathcal{G}$ through a directed path. A directed tree is an acyclic digraph with a root node, such that any other node of the digraph can be reached by one and only one directed path starting at the root node. A graph is said to have a spanning tree if a subset of the edges forms a directed tree.\par
Throughout the paper, $\lambda(.)$  represents the eigenvalues of a matrix. $(.)^{adj}$ refers to adjoint of a matrix. $ker(.)$ denotes the null space.  Furthermore, $\lambda_{max}(.)$ and $\lambda_{min}(.)$ represent maximum and minimum eigenvalue  of matrix, respectively. $diag(.)$ denotes the diagonal matrix. $a \otimes b$ represents Kronecker product of $a$ and $b$.

\smallskip

\noindent
$\textbf{Assumption 1.}$ The directed graph $\mathcal{G}$ has a spanning tree.   

\vspace{-0.2cm}

\subsection{Standard Distributed Consensus in  MAS}
This subsection presents the standard distributed control protocol for consensus of discrete-time MAS.

Consider \textit{N} agents with identical system dynamics represented by
\vspace{-0.3cm}
\begin{equation}
{x_i}(k + 1) = A{x_i}(k) + B{u_i}(k) ,\,\,\,\,\,\,\,i = 1,\dots,N
\label{eq_1}
\end{equation}

\vspace{-0.2cm}
\noindent
where ${x_i}(k) \in {R^{n}}$ and  ${u_i}(k) \in {R^{m}}$ are the state  and control input of agent $i$, respectively. $A$ and $B$ are the system and input matrices, respectively. $(A,B)$ is assumed to be stabilizable. 

Define  the local neighborhood tracking error for the agent $i$ as

\vspace{-0.5cm}

\begin{equation}
\varepsilon _i(k) = (1+ h_i)^{-1}\sum\limits_{j =1}^N {{a_{ij}}({x_j}(k) - {x_i}(k))} 
\label{eq_2}
\end{equation}

\vspace{-0.2cm}
\noindent
where $a_{ij}$ is the $(i,j)$-th value of the adjacency matrix.\par
Consider the distributed control law for each node $i$ as in \cite{c22}
\begin{equation}
u_{i}(k)=cK\varepsilon _i(k),\,\,\,\,\,\,\,i =1,\dots,N
\label{eq_3}
\end{equation}
where $c$ is a positive coupling gain and $K\in {R^{m \times n}}$ is a control gain, designed to gaurantee that agents reach consensus, i.e., $x_i(k) \to x_j(k) \,\,\forall i,j.$
Define the global state vector as $x(k)=[x_{1}^{T}(k), \,\,x_{2}^{T}(k),\,\dots,\,x_{N}^{T}(k)]^T \in R^{nN}$. Using (\ref{eq_1})-(\ref{eq_3}), the global dynamics of DMAS can be expressed as
\begin{align}
\begin{gathered}
{x}(k + 1) = [I_N \otimes A -c(I + H)^{-1}L\otimes BK]{x}(k)\\
\end{gathered}
\label{eq_4}
\end{align}

The normalized graph Laplacian matrix $\hat{L}$ is defined as \cite{c22}
\begin{equation}
\hat{L} =(I+ H)^{-1}L
\label{eq_41}
\end{equation}
Let the eigenvalues of the normalized graph Laplacian matrix $\hat{L}$ be $\lambda_i$, $\forall$ $i=1,\dots,N$. Then, $\lambda_i$ lies inside unit circle centered at $1+j0$ for $i=2,\dots,N$ and $\lambda_1=0$ \cite{c25}.

Using (\ref{eq_4}), the state of agent's global dynamics is given by
\begin{equation}
x(k) = [I_N \otimes A -c\hat{L}\otimes BK]^kx(0) \triangleq A_c^{k}x(0)
\label{eq_5a}
\end{equation}
where $A_c$ is the closed-loop matrix defined as  
\begin{equation}
A_c = (I_N \otimes A -c\hat{L}\otimes BK)
\label{eq_5}
\end{equation}

\noindent
\textbf{Lemma 1.} \cite{c25} Let $R \subset {\mathcal{V}}$ be the set of root nodes and $r = [{p_1},\dots,{p_N}]^{T}$ be the left eigenvector of the normalized graph Laplacian matrix $\hat{L}$ for $\lambda_1=0$. Then, ${p_i} > 0$ if $\,\,i \in R$ and ${p_i} = 0$ if $\,\,i \notin R$.
\smallskip

\noindent
\textbf{Theorem 1.} \cite{c22}-\cite{c25} 
Let feedback gain $K$  be designed such that $A-c\lambda_{i}BK$  is Schur stable for $i=2,\dots,N$. Then, according to Lemma $1$, the final consensus value for DMAS can  be written as
\begin{equation}
x(k) = ({r^T \otimes A^k})\left[ \begin{matrix}
{{x_1}(0)}  \\ 
   .  \\ 
   .  \\ 
   {{x_N}(0)}  \\ 
\end{matrix} \right]\,\,i=1,\dots,N\,\,\,as\,\,\,k \to \infty
\label{eq_6}
\end{equation}


\section{Attack Analysis for Discrete-time DMAS}
This section presents the attack modeling and analyzes its adverse effects on the standard control protocol. The internal model principle for the attacker is presented to show how a single compromised agent can destabilize the entire system. Then, the effect of the attack on the local neighborhood tracking error is analyzed to show the ineffectiveness of the standard robust $H_{\infty}$ control protocol (which is a well-known disturbance attenuation technique) in the presence of a stealthy attack.\par


Attacks on actuators of agent $i$ can be modeled as
\begin{equation}
u_{i}^{c}(k)=u_{i}(k)+\gamma_{i}u_{i}^{a}(k)
\label{eq_7}
\end{equation}
where $u_{i}$ is the control law given in (3), $u_{i}^{a}$ represents the attacker's signal injected into the actuator of agent $i$, $u_{i}^{c}$ is the distorted control law applied to (1) and the scalar $\gamma_{i}$  is 1 when there is an attack on actuators of agent $i$ and $0$, otherwise.\par
Attacks on sensors of agent $i$ can be modeled as
\begin{equation}
x_{i}^{c}(k)=x_{i}(k)+\delta_{i}x_{i}^{a}(k)
\label{eq_8}
\end{equation}
where $x_{i}$ represents the state of agent $i$, $x_{i}^{a}$ is the attacker's signal injected into the sensor of agent $i$, $x_{i}^{c}$ is the distorted state and the scalar $\delta_{i}$  is 1 when there is an attack on sensors of agent $i$ and $0$, otherwise.\par
Based on the distributed control law (\ref{eq_3}), and using (\ref{eq_7}) and (\ref{eq_8})  in  (\ref{eq_1}), one can express the DMAS dynamics for an agent $i$ as
\begin{equation}
{x_i}(k + 1) = A{x_i}(k) + B{u_i}(k) + Bf_i(k) ,\,\,\,\,\,\,\,i = 1,\dots,N
\label{eq_9}
\end{equation}
where $f_i(k)$ represents the overall attack signal injected into the agent $i$, which is given by
\begin{align}
\begin{gathered}
f_i(k) =  c(1+h_i)^{-1}K(\sum\limits_{j =1}^N {{a_{ij}}({\delta_{j}x_j^a}(k) - \delta_{i}{x_i^a}(k))} \\ 
+ \gamma_{i} {u_{i}^a}(k)
\end{gathered}
\label{eq_10}
\end{align}
\textbf{Remark 1.} An attacker  can manipulate sensors or actuators without physical tampering. Spoofing of global positioning system (GPS) of an unmanned vehicle or of phasor measurement unit's (PMU's) in power system are examples of attacks without physical tampering.   

\subsection{Effects of Attack on Standard DMAS}
This subsection analyzes the effects of the attack on the standard discrete-time DMAS (\ref{eq_1}). Theorem $2$ investigates how an attack can propagate across the network.
\smallskip

\noindent
\textbf{Definition 1.} In a graph, agent $i$ is \textit{reachable} from agent $j$ if there is a directed path of any length from node $j$ to node $i$.
\smallskip

\noindent
\textbf{Definition 2.}  An agent is said to be a \textit{compromised} agent, if it is directly affected by the attacker.
\smallskip

\noindent
\textbf{Theorem 2.} Consider the discrete-time DMAS (\ref{eq_9}) under the attack $f_i(k)$. Let the control protocol be designed as (\ref{eq_3}) such that the closed loop matrix $A_c$ in (\ref{eq_5}) is Schur. Then, 
\begin{enumerate}
\item All agents reach consensus if $f_i(k)=0, \forall i=1,\dots,N.$
\item The intact agent deviates from the desired consensus value if it is reachable from a compromised agent.
\item The deviation of the network from the desired behavior depends on the number of compromised agents, their attack signal magnitude and the number of agents reachable from them.
\end{enumerate}

\smallskip


\noindent
$\textit{Proof.}$  It is shown in \cite{c25} that if $c$ and $K$ are designed so that $A_c$ in (\ref{eq_5}) is Schur, then all agents reach consensus. This completes the proof of part $1$.\par
To prove part $2$, define ${x^a(k)} = [{(x_1^a(k))^T},{(x_2^a(k))^T},\dots,{(x_N^a(k))^T}]^T$ and ${u^a(k)} = [{(u_1^a(k))^T},{(u_2^a(k))^T},\dots,{(u_N^a(k))^T}]^T$ as a vector of signals injected to the sensors and actuators, respectively. The global dynamics for DMAS (\ref{eq_9}) under the effect of  attack can be written as
\begin{equation}
{x}(k + 1) = A_c{x}(k) +  (I_N \otimes B)f(k) ,\,\,\,\,\,\,\,i = 1,\dots,N
\label{eq_13}
\end{equation}
where the injected global attack signal $f(k)$ is 
\begin{equation}
f(k) =  - c(\hat{L} \otimes K)(\delta \otimes I_N){x^a} + (\gamma \otimes I_N){u^a}
\label{eq_14}
\end{equation}
with  $\gamma  = diag({\gamma _1},\dots,{\gamma _N}) $ and $\delta  = diag({\delta_1},\dots,{\delta_N}) $. If  $f(k) \ne 0$, then the solution of (\ref{eq_13}) is given by
\begin{equation}
x(k) = A_c^{k}x(0)\, + \,\sum\limits_{p = 0}^{k - 1} {(A_c)^{k - p - 1}(I_N \otimes B)f(p)} 
\label{eq_15}
\end{equation}
with $k\geq p$.
Then, one can write (\ref{eq_15}) as
\begin{align}
\begin{gathered}
  x(k) = A_c^{k}x(0)\, + \sum\limits_{p = 0}^{k - 1} {{{({I_N} \otimes A)}^{k - p - 1}}({I_N} \otimes {I_n}} \,\,\,\, \hfill \\
  \,\,\,\,\,\,\,\,\,\,\,\,\,\,\,\,\,\,\,\, - c\hat L \otimes {A^{ - 1}}BK{)^{k - p - 1}}({I_N} \otimes B)f(p) \hfill \\ 
\end{gathered} 
\label{eq_16}
\end{align}
in which the second part of (\ref{eq_16}) reflects the effect of the attackers' input on the system. For a positive integer $n$, the binomial theorem for matrices can be expressed as ${(x + y)^n} = \sum\limits_{k = 0}^n {C_k^n} {x^{n - k}}{y^k}$ with ${C_k^n}=\frac{n!}{(n-k)!k!}$ if $x$ and $y$ is commutative. Using this fact and Theorem 1, (\ref{eq_16}) becomes
\begin{align}
\begin{gathered}
   x(k) = ({r^T \otimes A^k})x(0)\, + \sum\limits_{p = 0}^{k - 1} {{{({I_N} \otimes A)}^{k - p - 1}}} \,\,\,\, \hfill \\
  \,\,\,\,\,\,\,\,\,\,\,\,\,\,\,\,\,\,*\sum\limits_{m = 0}^{k - p - 1} {C_m^{k - p - 1}} {( - c\hat L \otimes {A^{ - 1}}BK)^m}({I_N} \otimes B)f(p) \hfill \\ 
\end{gathered}
\label{eq_18}
\end{align}
Using (\ref{eq_18}), the state of the agent $i$ at steady state can be written as
\begin{align}
\begin{gathered}
x_i(k)  \to r^TA^Kx(0)+ \hfill \\ \sum\limits_{j=1}^{N} \sum\limits_{p = 0}^{k - 1} {\sum\limits_{m = 0}^{k - p-1} A^{k-p-1}C_m^{k-p-1}(-1)^mc^ml{_{ij}^m}({A^{-1}BK})^{m}Bf_j(p)}
\label{eq_19}
  \end{gathered}
\end{align}
in which first term represents the desired consensus value that depends on the eigenvalues of the system dynamics matrix $A$ (i.e., consensus value becomes zero for the stable system dynamics $A$ or non-zero for the marginally stable system dynamics $A$). $l_{ij}^m \triangleq [(I+ H)^{-1}L]_{ij}^m$  with $[\,\,]_{ij}$ denotes the element $(i,j)$ of a matrix. $m$ represents the length of shortest directed path from $j$ to $i$  \cite{vad}. Assume now that the agent $j$ is under direct attack, but agent $i$ is  intact, i.e. $f_i(k)=0$ and $f_j(k) \neq 0$. If the intact agent $i$ is reachable from the compromised agent $j$, since $l_{ij}^m \ne 0$ for some $0<m<N-1$, one can infer from (\ref{eq_19}) that  the agent state $x_i(k)$ at steady state in (\ref{eq_19}) has nonzero second part which deduces that intact agent $i$ is deviated from the desired consensus behavior.  This completes the proof of part $2$. \par

For the proof of part $3$, taking the norm from both sides of (\ref{eq_15}) yields 
\begin{equation}
\left\| {x(k)} \right\|   \leqslant  \left\|A_c^{k}x(0)\right\| + \, \left\|\sum\limits_{p = 0}^{k - 1} {(A_c)^{k - p - 1}(I_N \otimes B)f(p)}\right\| 
\label{eq_15_a}
\end{equation} 
and using 
\begin{equation}
\left\| {\sum\limits_{p = 0}^{k - 1} {{{({A_c})}^{k - p - 1}}f(p)} } \right\| \leqslant \frac{{\left\| {f(k)} \right\|}}{{\left| {{\lambda _{\min }}({A_c})} \right|}}
\end{equation}
one can write (\ref{eq_15}) 
\begin{equation}
\left\| {x(k)} \right\|   \leqslant  \left\|({r^T \otimes A^k})x(0)\right\| + \, {N_f}\frac{{\left\| B \right\|{b_f}}}{{\left| {{\lambda _{\min }}({A_c})} \right|}}
\label{eq_15_b}
\end{equation} 
at steady state, where, $N_f$ is the number of agents for which $f_i(k)$ is non-zero, $b_f$ is bound on adversarial input $f_i(k)$. It was shown in part $2$ that if agent $i$ is reachable from the compromised agent $j$, then its deviation from the desired behavior is nonzero. That is, for the agent $i$ which is reachable from a compromised agent, the deviation in $\left\| {x_i(k)} \right\|$ in (\ref{eq_15_b}) depends on the number of compromised agents  $N_f$  and bound on adversarial input $b_f$. This completes the proof. \hfill $\square$

\subsection{Internal Model Principle Approach for the Attacker}
In the control systems, to reject a disturbance or follow a reference trajectory, one needs to incorporate reference dynamics in the system. This is called the internal model principle (IMP). We showed in the following Theorem $3$ that the attacker can also leverage the IMP and incorporate some eigenvalues of the consensus dynamics in its attack design to destabilize the entire network. \par

We now take the role of the attacker and show that how it can maximize the damage and cause a catastrophe. Conditions under which the attacker achieves this objective are provided. 

\smallskip


\noindent
\textbf{Definition 3.} (\textit{IMP-based and non-IMP-based Attacks}.) Let the attack signal $f_i(k)$  be generated by
\begin{equation}
f_i(k+1) = Wf_i(k)
\label{eq_20}
\end{equation}
where ${W} \in R^{m \times m}$ denotes the dynamics of the attack signal. Define  
\begin{equation}
\left\{ \begin{array}{l}
{\Lambda _W} = [{\lambda _{{W_1}}}, \ldots ,{\lambda _{{W_m}}}]\\
{\Lambda _A} = [{\lambda _{{A_1}}}, \ldots ,{\lambda _{{A_n}}}]
\end{array} \right.
\label{eq_20a}
\end{equation}
as the set of eigenvalues of the attack dynamics ${W}$  and the system dynamics matrix $A$, respectively.
Then, if $\Lambda_{{W}} \subseteq \Lambda_{A}$, the attack signal is called the IMP-based attack. Otherwise, if $\Lambda_{{W}} \not\subset \Lambda_{A}$ or  the attacker has no dynamics (e.g. a random signal), it is called a non-IMP based attack.\par

We assume that the system matrix $A$ in (1) is marginally stable, with  eigenvalues on the unit circle centered at origin. This is a standard assumption in the literature for consensus and synchronization problems \cite{jhg}. In fact, if $A$ has stable eigenvalues, one can ignore them and reduce the dimension of $A$. This is because, stable states of the agent have no effect on the steady-state synchronization trajectory, and only contribute to the transient response.Define 
\begin{equation}
{S}(k) = \sum\nolimits_{j = 1}^N {{p_{1j}}{f_j(k)}}
\label{eq_20c}
\end{equation}
where ${p_{1j}}$ represents the element of left eigenvector corresponding to zero eigenvalue of $\hat{L}$. Based on Lemma 1, one can conclude that $S(k) \ne 0$ if  $j \in R$, i.e., if attack is on a root node, and $S(k)=0$ otherwise.





\smallskip

\noindent
\noindent
\textbf{Theorem 3.} Consider the DMAS (\ref{eq_9}) under the attack $f_i(k)$ with the control protocol (\ref{eq_3}). Let ${f_i}(k)$ be designed as (\ref{eq_20}). Then, 

\begin{enumerate}
  \item An IMP-based attack destabilizes the complete network, if ${S(k) \ne 0}$, i.e., if attack is on a root node.

\item Any non-IMP based attack or IMP-based attack with ${S(k) = 0}$ deviates agents from the desired consensus behavior but does not cause instability, if agents are reachable from the compromised one. 
\end{enumerate}

\vspace{0.25cm}
\textit{Proof.}
The transfer function for the DMAS (\ref{eq_1}), from ${x_i}(z)$ to ${u_i}(z)$ in z-domain can be written as
\begin{equation}
{G_i}(z) = \frac{{{x_i}(z)\,\,}}{{{u_i}(z)}} = (zI-A)^{-1}B
\label{Im_3}
\end{equation}

Using (\ref{eq_3}), the global control law under the influence of the attack can be expressed as
\begin{equation}
u(z) =-(c\hat{L} \otimes K)x(z) + f(z)
\label{Im_4}
\end{equation}
with $u(z) = {[u_{_1}^T,\dots,u_{_N}^T]^T}$, $\,$  $x(z) = {[x_{_1}^T,\dots,x_{_N}^T]^T}$ and $f(z) = {[f_{_1}^T,\dots,f_{_N}^T]^T}$. Using (\ref{Im_3}) and (\ref{Im_4}), the system state in the global form can be written as
\begin{equation}
x(z) = (I_N \otimes G(z))u(z) = (I_N \otimes G(z))(-(c\hat{L} \otimes K)x(z) + f(z)) 
\label{Im_5}
\end{equation}
where $G(z)=diag(G_{i}(z))$ with dimension ${R^{NxN}}$. Let $M$ be a non-singular matrix such that $\hat{L}=M\Lambda {M^{-1}}$, with $\Lambda$ be the Jordan canonical form of the normalized graph Laplacian matrix $\hat{L}$. The left and the right eigenvectors of $\hat{L}$ corresponding to the zero eigenvalue of the normalized graph Laplacian matrix are $r$ and $\mathbf{1}_N$, respectively \cite{c25}. Define
$$M = [1\,\,\,{M_1}],\hspace{0.5cm}{M^{ - 1}} = [{{r^T}}\,\,{{M_2}}]^T$$
where  ${M_1} \in {R^{N \times (N - 1)}}$ and ${M_2} \in {R^{(N - 1) \times N}}$. Using (\ref{Im_5}) with $\hat{L}=M\Lambda {M^{-1}}$, one has
\begin{align}
[I_{Nn} + cM\Lambda {M^{ - 1}} \otimes G(z)K]x(z) = (I_N \otimes G(z))f(z)
\label{Im_6}
\end{align}
As $M{M^{ - 1}} = {I_N}$, one can write (\ref{Im_6}) as 
\begin{align}
(M \otimes I_n)[{I_{nN}} + c\Lambda \otimes G(z)K ]({M^{ - 1}} \otimes I_n)x(z) = G(z)f(z)
\label{Im_6_1}
\end{align}
Defining a state transformation as
\begin{equation}
\hat{x}(z)=({M^{- 1}} \otimes I_n)x(z)
\label{Im_6_2}
\end{equation}
and premultiplying (\ref{Im_6_1}) with $({M^{- 1}} \otimes I_n)$ gives
\begin{equation}
 \hat{x}(z)= [{I_{Nn}} + c\Lambda \otimes G(z)K ]^{-1}({M^{- 1}} \otimes G(z))f(z)
\label{Im_8}
\end{equation}
Let assume for simplicity that all the Jordan blocks are simple, $M^{-1}=[p_{ij}]$ and $M=[m_{ij}]$, where ${p_{ij}}$ and $m_{ij}$ represent the elements of matrices $M^{-1}$ and $M$,  formed by left eigenvectors and right eigenvectors of the normalized graph Laplacian matrix $\hat{L}$, respectively.  For the agent $i$, using (\ref{Im_6_2}) and (\ref{Im_8}), one has
\begin{equation}
\begin{gathered}
{x_i}(z)={\sum\limits_{h=1}^N {m_{ih}[I_n + cK{G_i}(z)\lambda_i]^{-1}{G_i}(z){\sum\nolimits_{j = 1}^N {p_{ij}}{f_j}(z)}}}
\end{gathered}
\label{Im_9}
\end{equation}
The first eigenvalue of the normalized graph Laplacian matrix $\hat{L}$ is zero and its corresponding right eigenvector is $\mathbf{1}_N$ i.e. $m_{i1}=1$. Using this fact with (\ref{Im_9}), one has 
\begin{align}
\begin{gathered}
  {x_i}(z) = {G_i}(z)\sum\limits_{j = 1}^N {{p_{1j}}} {f_j}(z) +  \hfill \\
  \sum\limits_{h = 2}^N {{m_{ih}}} {[I_n + c{\lambda _h}{G_i}(z)K]^{ - 1}}{G_i}(z)\sum\limits_{j = 1}^N {{p_{hj}}} {f_j}(z) \hfill 
 \label{Im_11}
\end{gathered} 
\end{align}
 Now, if we show that $[I_n +  cK{G_i}(z)\lambda_h]^{ - 1}$ is Schur, then the second term of (\ref{Im_11}) is bounded, even in the presence of attack.\par
Since $(A - c{\lambda _h}BK),\,\,\forall h = 2,\dots,N$ is Schur,  therefore if we show that the roots of the characteristic polynomial $(A - c{\lambda _h}BK)$ are  identical to the poles of $[I_n +  cK{G_i}(z)\lambda_h]^{ - 1}$, then one can say $[I_n +  cK{G_i}(z)\lambda_h]^{ - 1}$ is also Schur. To this end, using (\ref{Im_3}), one has
\begin{align}
\begin{gathered}
  \Delta |(z{I_n} - (A - c{\lambda _h}BK))| = \Delta |(z{I_n} - A + c{\lambda _h}BK)| \hfill \\
   = \Delta |z{I_n} - A|({I_n} + c{\lambda _h}{(z{I_n} - A)^{ - 1}}BK) \hfill \\
   = \frac{{\Delta |z{I_n} - A|[(\Delta |z{I_n} - A| + c{\lambda _h}{{(z{I_n} - A)}^{adj}}BK)]}}{{\Delta |z{I_n} - A|}} \hfill \\ 
\end{gathered} 
\label{Im_12}
\end{align} 
Hence, this proves that the roots of the characteristic polynomial $(A - c{\lambda _h}BK)$  are  identical to the poles of $[I_n +  cK{G_i}(z)\lambda_h]^{ - 1}$ using matrix properties from \cite{c27}. Therefore, $[I_n +  cK{G_i}(z)\lambda_h]^{ - 1}$ is Schur. Thus, it concludes that the second term of (\ref{Im_11}) is bounded and has no contribution in destabilizing the system. \par

\smallskip

According to Lemma 1, $\sum\nolimits_{j = 1}^N {p_{1j}}{f_j}(k)$ or $S(k)$ in (\ref{eq_20c}) is zero for attack on non-root nodes and  nonzero, if the attack is launched on root nodes. Consider an IMP-based attack on a root node. Then, using the transfer function  (\ref{Im_3}) and the attack signal defined in (\ref{eq_20}), one can write (\ref{Im_11}) as
\begin{align}
\begin{gathered}
  {x_i}(z) = \sum\limits_{j = 1}^N {{p_{1j}}\frac{{{{(z{I_n} - A)}^{adj}}B{{(z{I_n} - W)}^{adj}}{f_i}(0)}}{{{{({z^2} + \lambda _{{A_l}}^2)}^2}\{ \prod\limits_{i = 1,i \ne l}^n {({z^2} + \lambda _{{A_i}}^2){{({z^2} + \lambda _{{W_i}}^2)}}} \} }}} +  \hfill \\
  \sum\limits_{h = 2}^N {{m_{ih}}} {[1 + cK{G_i}(z){\lambda _h}]^{ - 1}}{G_i}(z)\sum\limits_{j = 1}^N {{p_{hj}}} {f_j}(z) \hfill \\ 
\end{gathered} 
\label{Im_14}
\end{align}
The first term of (\ref{Im_14}) shows that the pole $\lambda _{{A_l}}$ lies on the unit circle centered at the origin and has multiplicity greater than 1. Thus, the system states tend to infinity in the discrete-time domain as $k \to \infty$. Therefore, the attack on the root node destabilizes the entire network.  This completes the proof of part $1$.\par
If the attack is on a non-root node, then $\sum\nolimits_{j = 1}^N {{p_{1j}}{f_j}(k) = 0}.$ So,  (\ref{Im_11}) can be expressed as 
\begin{equation}
{x_i}(z) = \sum\limits_{h = 2}^N {{m_{ih}}} {[1 +  cK{G_i}(z)\lambda_h]^{ - 1}}{G_i}(z)\sum\limits_{j = 1}^N {{p_{hj}}} {f_j}(z)
\label{Im_15}
\end{equation}
Then, according to (\ref{Im_12}),   $[I_n +  cK{G_i}(z)\lambda_h]^{ - 1}$  is Schur stable. Therefore, the system states are bounded, even in the presence of the attack. Moreover, the agents that are reachable from the attacker shows stable behavior, but deviation from the desired consensus value. 
If $\Lambda_{A} \cap \Lambda_{W} \ne \phi $  which implies that the multiplicity of poles lie on the unit is one. Therefore according to  (\ref{Im_11}), the system states remain bounded and shows deviation from the desired consensus behavior due to the adverse effect of the attacker. This completes the proof. $\hfill$  $\square$

\smallskip

\noindent
\textbf{Remark 2.} Note that the attacker does not need to know the system matrix $A$, and it can identify the eigenvalues of dynamics through eavesdropping the sensory informations. Then, the attacker can identify root node and destabilize the entire system.

\smallskip
The following example presents the adverse effect of attack on the root node.

\medskip

\noindent
\textbf{Example 1.}  Consider $4$ agents having single-integrator dynamics  given by 
 \begin{equation}
{x_i}(k + 1) = {x_i}(k) + {u_i}(k) \,\,\, i=1,\dots,4
\label{ex1}
 \end{equation}
with the control protocol (\ref{eq_3}) and communicating to each other according to graph structure in Fig.1. \par

In the absence of attack signal, agents achieve the desired consensus value,  which is the average of the initial values of Agents $1$ and $2$ in this example. If the attacker launches an attack on the Agent $1$, then, the dynamics of the system at the steady state can be written as
\begin{equation}
\begin{gathered}
  {x_1}(k + 1) = {x_1}(k) + {u_1}(k) + {u_{ac}}(k) = 0 \hfill \\ 
  \Rightarrow {x_1}(k) + {(1 + {h_1})^{ - 1}}({x_2}(k) - {x_1}(k)) + 1 = 0\hfill \\
  \Rightarrow ({x_1}(k) + {x_2}(k)) =  - 2 \hfill  \\ 
  \end{gathered}
\label{ex2a}
\end{equation}
\begin{equation}
\begin{gathered}
  {x_2}(k + 1) = {x_2}(k) + {u_2}(k) = 0\hfill \\
  \Rightarrow \,\,\,{x_2}(k) + {(1 + {h_2})^{ - 1}}({x_1}(k) - {x_2}(k)) = 0\hfill \\
  \Rightarrow ({x_1}(k) + {x_2}(k)) = 0 \hfill  \\
  \end{gathered}
\label{ex2b}
\end{equation}
\vspace{-0.5cm}
\begin{figure}[H]
\begin{center}
\includegraphics[width=80mm,height=45mm]{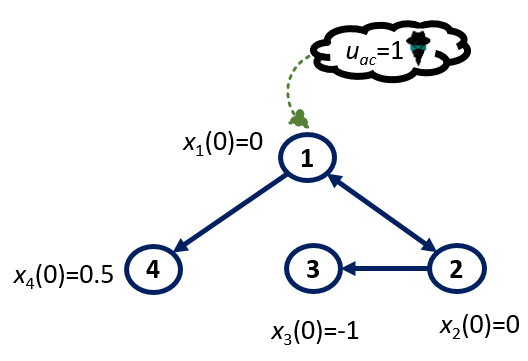}
\end{center}
\caption{Graph topology}
\label{1}
\end{figure}

\begin{equation}
\begin{gathered}
  {x_3}(k + 1) = {x_3}(k) + {u_3}(k) = 0\hfill \\
  \Rightarrow \,\,\,{x_3}(k) + {(1 + {h_3})^{ - 1}}({x_2}(k) - {x_3}(k)) =0\hfill \\
  \Rightarrow \,\,({x_2}(k) + {x_3}(k)) = 0 \hfill  \\
  \end{gathered}
\label{ex2c}
\end{equation}
\begin{equation}
\begin{gathered}
  {x_4}(k + 1) = {x_4}(k) + {u_4}(k) = 0\hfill  \\
\Rightarrow \,\,{x_4}(k) + {(1 + {h_4})^{ - 1}}({x_1}(k) - {x_4}(k)) = 0\hfill \\ 
\Rightarrow ({x_1}(k) + {x_4}(k)) = 0 \hfill   \\ 
\end{gathered}
\label{ex2d}
\end{equation}
However, the dynamics in (\ref{ex2a}) and (\ref{ex2b}) show that to reach a steady state, $({x_1}(k) + {x_2}(k)) = - 2$ and  $({x_1}(k) + {x_2}(k)) = 0$ at same time. This is not possible for consensus on a bounded state, and this can happen only, if they both go to infinity. Therefore, the system states never achieve the desired consensus behavior and they converge to infinity. \par

When Agent $1$ is attacked with an IMP-based adversarial input $u_{ac}=1$, using Z-transform, one has
\begin{equation}
(z - 1){x_1}(z) = {u_1}(z) + {u_{ac}}(z)\, \Rightarrow {x_1}(z) = \frac{{{u_1}(z)}}{{(z - 1)}} + \frac{z}{{{{(z - 1)}^2}}}
\label{ex31}
\end{equation}
It can be seen from (\ref{ex31}) that an IMP-based attack can destabilize the entire system. This verifies the results of Theorem $3$.

\smallskip

Now, we present the analysis of the effects of the attack on the local neighborhood tracking error (\ref{eq_2}). This analysis shows that although attacks on sensors 
and actuators can be modeled as disturbances, existing disturbance attenuation techniques do not work for attack attenuation.

Disturbance attenuation approaches focus on minimizing the effects of disturbance on the local neighborhood tracking error \cite{c23}-\cite{c24}. More specifically, the $H_{\infty}$ approach for DMAS (\ref{eq_1}) in presence of disturbance $w_i(k)$  designs a distributed control protocol as in (\ref{eq_3}), such that the  desired  consensus is achieved as in (\ref{eq_6}), if disturbance $w_i(k)=0$ and the bounded $L_2$-gain condition is fulfilled for any disturbance $w_i(k) \in L_2[0,\infty)$
\begin{equation}
\sum\limits_{k = 0}^\infty  {{\varepsilon ^T}(k)\bar{M}\varepsilon } (k) \leqslant {\gamma ^2}\sum\limits_{k = 0}^\infty  {{w^T}(k)\bar{N}w(k)} 
\label{hin}
\end{equation}
where $\gamma>0$ is attenuation constant, $\bar{M}$ and $\bar{N}$ are positive definite weight matrices.\par

We present the following rigorous analysis for the effects of the attack on the local neighborhood tracking error in following Theorem $4$ and show that how an attacker can bypass existing $H_{\infty}$ disturbance attenuation approaches and make them entirely ineffective.\par

\smallskip

\noindent
\textbf{Lemma 2.} Consider the normalized graph Laplacian matrix $\hat L$ defined in  (\ref{eq_41}).  Then, $[{\hat{L}^T}{\hat{L}} - 2\hat{L}]$ is  negative semidefinite.

\smallskip

\noindent
\textit{Proof.} Let $\lambda_k$ be the eigenvalue of the normalized graph Laplacian matrix ${\hat{L}}$. So, the eigenvalue of $[{\hat{L}^T}{\hat{L}} - 2\hat{L}]$  can be written as
\begin{align}
\nonumber
eig[{\hat{L}^T}{\hat{L}} - 2\hat{L}]= \lambda _k^2 - 2{\lambda _k}\\
=({\lambda _k} - 1)^2- 1
\label{Ad_1}
\end{align}

Since all eigenvalues of matrix ${\hat{L}}$ lie inside unit circle centered at $1+j0$, except ${\lambda _1}=0$  \cite{c25}, therefore $({\lambda _k} - 1)^2- 1$ is less than or equal to zero for $k=1,\dots,N$. This shows that $[{\hat{L}^T}{\hat{L}} - 2\hat{L}]$ is negative semidefinite. 
\smallskip

In the following theorem, for the sake of simplicity, we consider the single integrator dynamics (\ref{ex1}) and its global dynamics is given by
\begin{align}
x(k+1)= x(k) + u(k)
\label{si_1}
\end{align}
Under the influence of attack, one can write the control input $u(k)$ in (\ref{si_1}) as 
\begin{align}
u(k)=(-\hat{L}x(k) + f(k))
\label{si_2}
\end{align}

\noindent
\textbf{Theorem 4.} Consider the discrete-time DMAS with single integrator dynamics (\ref{si_1}). Assume that the system is under a constant attack signal $f(k)$. Then, the local neighborhood tracking error for intact agents is zero while agents do not reach the desired consensus.

\smallskip

\noindent
\textit{Proof.}  Consider the Lyapunov function for the discrete-time DMAS as 
\begin{equation}
V(x(k),f(k)) = {(-\hat{L}x(k) + f(k))^T}{(-{\hat{L}}x(k) + f(k))}
\label{Im_16}
\end{equation}
The difference equation of the Lyapunov function (\ref{Im_16}) can be written as
\begin{equation}
\nonumber \Delta V(x(k),f(k))= V(x(k + 1),f(k + 1)) - V(x(k),f(k))
\end{equation}
\begin{align}
 = \nonumber ( - \hat{L}x(k + 1) + f(k + 1))^T( -\hat{L}x(k + 1) + f(k + 1)) \\ - ( - \hat{L}x(k) + f(k))^T( -\hat{L}x(k) + f(k))
\label{Im_17}
\end{align}
For the constant attack signal $f(k+1)=f(k)$, one can write (\ref{Im_17}) as
\begin{align}
\nonumber = {( - \hat{L}x(k + 1) + f(k))^T}( - \hat{L}x(k + 1) + f(k)) \\ \nonumber - {( - \hat{L}x(k) + f(k))^T}( - \hat{L}x(k) + f(k)) 
\end{align}
or equivalently,
\begin{align}
\nonumber = {( -\hat{L}x(k + 1))^T}( -\hat{L}x(k + 1)) - {( - \hat{L}x(k))^T}(-\hat{L}x(k)) \\- 2f{(k)^T}\hat{L}(x(k + 1) - x(k))
\label{Im_18}
\end{align}
Using the scalar system dynamics (\ref{si_1}) in (\ref{Im_18}), one has
\begin{align}
\nonumber = {( - \hat{L}[x(k) + u(k)])^T}( - \hat{L}[x(k) + u(k)]) \\- {( - \hat{L}x(k))^T}( - \hat{L}x(k)) - 2f{(k)^T}\hat{L}u(k)
\label{Im_19}
\end{align} 
Using (\ref{si_2}), equation (\ref{Im_19}) can be written as
\begin{align}
\nonumber = {( - \hat{L}[x(k) -\hat{L}x(k) + f(k))])^T}( - \hat{L}[x(k) -\hat{L}x(k) + f(k))]) \\- {( - \hat{L}x(k))^T}( - \hat{L}x(k)) - 2f{(k)^T}\hat{L}(-\hat{L}x(k) + f(k)))
\label{Im_19a}
\end{align}
one can further simplify (\ref{Im_19a}) as
\begin{align}
= {( - \hat{L}x(k) + f(k))^T}[{\hat{L}^T}\hat{L} - 2\hat{L}]( -\hat{L}x(k) + f(k)) 
\label{Im_20}
\end{align}
Using Lemma $2$, one has
\begin{align}
\nonumber \Delta V(x(k),f(k)) = (-\hat{L}x(k) + f(k))^T[\hat{L}^T\hat{L} \\ - 2\hat{L}]( -\hat{L}x(k) + f(k)) \leqslant 0
\label{Im_21}
\end{align}
Then, using Lasalle's invariance principle \cite{c26}, the trajectories $(x(k),f(k))$ converge to a set that satisfy $\Delta V(x(k),f(k)) = 0$. Based on (\ref{Im_21}), this yields
\begin{align}
(-\hat{L}x(k) + f(k))  \in \ker (\hat{L}^T\hat{L} - 2\hat{L})\\
         \nonumber or \hspace{2cm} \hfill  \\
 (-\hat{L}x(k) + f(k)) = 0
\label{Im_22}
\end{align}
From (54), one has $(-\hat{L}x(k) + f(k)) = \bar c \mathbf{1}_N$. According to this, the single integrator system dynamics  becomes ${x_i}(k + 1) = {x_i}(k) + \bar c,$  which shows that it destabilizes the system. Therefore,  ${x_i}(k) \to \infty $ as $k \to \infty $ $\forall i=1,\dots,N$ with the local neighborhood tracking error goes to zero for all agent. Note that, based on Theorem $3$,  (54) is the possible case when the attack is on a root node. 
On the other hand, for an attack on a non-root node agent, from (\ref{Im_22}), one has $(-\hat{L}x(k) + f(k))=0$.
Since for the intact agent $i$, $f_i(k)=0$, therefore, the local neighborhood tracking error for intact agents converge to zero, even in the presence of the attack. 

We now show that intact agents do not reach the desired consensus, despite the fact the local neighborhood tracking error is zero. From (\ref{Im_22}), one has
\begin{equation}
\hat{L}x(k) =f(k)
\label{c11}
\end{equation}
which can be written for agent $i$ as
\begin{equation}
(1+ h_i)^{-1}\sum\limits_{j =1}^N {{a_{ij}}({x_j}(k) - {x_i}(k))} =f_i(k)
\label{c12}
\end{equation}
For a compromised agent $i$, since $f_i(k) \neq 0$, then, one has  $x_i(k) \neq x_j(k)$ for some $i,j$. \par
Now let assume that agent $i$ is intact. Then, one has 
\begin{equation}
(1+ h_i)^{-1}\sum\limits_{j =1}^N {{a_{ij}}({x_j}(k) - {x_i}(k))} =0
\label{c13}
\end{equation}
Consider the intact agent $i$ as an immediate neighbor of the compromised agent $i_c$. Let assume by contradiction that only the compromised agent does not reach the desired consensus but all the intact agents reach the desired consensus. Using (\ref{c13}), one can write
\begin{equation}
{(1 + {h_i})^{ - 1}}\sum\limits_{j \in {N_i}} {{a_{ij}}({x_j} - {x_i}) + } {a_{i{i_c}}}({x_{{i_c}}} - {x_i})=0
\label{c14}
\end{equation}
Assuming that intact agents reach consensus, $x_i(k) = x_j(k) \,\, \forall j \in {N_i}$. However, (\ref{c14}) cannot be satisfied if  $x_i(k) = x_j(k) \,\, \forall j \in {N_i}$ because $x_{i_c}(k) \neq x_i(k)$ and this contradict the assumption. Therefore, this shows that the intact agent $i$ is deviated from the desired consensus value. Similarly, one can use the same argument to show that all reachable agents from the compromised agent will deviate from the desired consensus value. This completes the proof.  \hfill $\square$

\smallskip

\noindent
\textbf{Remark 3.} If an intact agent $i$ is an immediate neighbor of a compromised agent $i_c$, then using (\ref{eq_2}),  one can write  the local neighborhood tracking error $\epsilon_i(k)$ with $a_{ij}=1$  as
\begin{equation}
\begin{gathered}
  {\varepsilon _i}(k) = {(1 + {h_i})^{ - 1}}\sum\limits_{j \in {N_i}} {({x_j}(k) - {x_i}(k))}  \hfill \\
  \,\,= \left| {{N_i}} \right|\,{(1 + {h_i})^{ - 1}}(\frac{1}{{\left| {{N_i}} \right|}}\sum\limits_{j \in {N_i}} {{x_j}(k)} \, - {x_i})\hfill \\ \,\,= \left| {{N_i}} \right|\,{(1 + {h_i})^{ - 1}}({x_{avg}}\, - {x_i}) \hfill \\ 
\end{gathered} 
\label{eq_2q}
\end{equation}
where ${x_{avg}}= \frac{{\sum\limits_{j \in {N_i}} {{x_j}(k)} }}{{\left| {{N_i}} \right|\,}}, $ which is not equal to $ x_i(k)$ due to incoming information from a comprised agent $ x_{ic}(k)$. From (\ref{eq_2q}), one can infer that the deviation of the intact agent from the desired consensus value depends on the number of the in-neighbors and deviation of the compromised agent $i_c$ from the desired consensus value which depends on the magnitude of the injected attack signal. Moreover, the closer the agent is to the source of the attack, the more its value will be deviated from the desired consensus.

\smallskip

\noindent
\textbf{Corollary 1.} Let the attacker design its attack signal using the internal model principle approach described in Theorem $3$. Then, it  bypasses the  $H_{\infty}$ control protocol.
\smallskip 

\noindent
\textit{Proof.} In the absence of the attack, minimizing the local neighborhood tracking error results in minimizing the consensus error. Therefore, the  $H_{\infty}$ control in (\ref{hin}) is used to attenuate the effect of adversarial input on the local neighborhood tracking error. However, according to Theorem $4$, in the presence of IMP attack, by making  the local neighborhood tracking error go to zero, agents do not reach consensus. This completes the proof. \hfill $\square$
\smallskip

\noindent

Theorem $4$ and the following analysis highlight that while the local neighborhood tracking error is zero, agents might not reach consensus. Now, define a global performance function $\Gamma (k)$ as 
\begin{equation}
\Gamma (k)  = \sum\limits_{i \in {N}}\sum\limits_{j \in {N_i}} {{\left\| {{x_i}(k) - {x_j}(k)} \right\|} ^2}
\label{d71}
\end{equation}
Define the set of intact agents as
\begin{equation}
N_{int}=N-N_c  
\end{equation}
where $N$ represents set of all agents and $N_c$ represents set of compromised agents in the network.

\smallskip

\noindent
\textbf{Corollary 2.}  Consider the  global performance function $\Gamma(k)$  and the local neighborhood tracking error $\epsilon_i(k)$ defined in (\ref{d71}) and (\ref{eq_3}), respectively. Then,  



\begin{enumerate}
\item  $\Gamma(k)$  and $\epsilon_i(k)$ $\forall i=1,\dots,N$ converge to zero, if there is no attack. Moreover, agents achieve the desired consensus.
\item  If the attacker designs an IMP-based attack on the non-root node, then $\epsilon_{i}(k) \,\,\,\,\,\,\,\forall i\in N_{int}$ converges to zero, but $\Gamma(k)$  does not converges to zero. That is, agents do not reach the desired consensus, while the local neighborhood tracking error is zero.
\item  If the attacker designs an IMP-based attack on the root node, then $\epsilon_i(k)$  and $\Gamma(k)$ $\forall i=1,\dots,N$ go to zero,  despite agents do not achieve the desired consensus and the entire system get destabilized. 
\end{enumerate}
\smallskip

\noindent
\textit{Proof.} According to Theorem $1$, the system achieves the desired consensus if there is no adversarial input in the system and this proves part $1$ of corollary. If the attacker injects an IMP-based attack signal into the non-root node of the DMAS, then based on Theorem, $3$ and $4$, one can infer $\epsilon_i(k) \to 0$. However, as shown in Theorem $4$, ${x_i}(k)-{x_j}(k)\not \to 0$, so  $\Gamma(k)\not \to 0$ and this proves part $2$. Based on Theorem $3$, if the attacker injects an IMP-based attack signal into the root node of the DMAS, then ${x_i}(k)-{x_j}(k)\to 0$. However, the system gets destabilized as ${x_i}(k) \to \infty$ as $k \to \infty$, while $\epsilon_i(k)$ and $\Gamma(k)$  $\,\,\,\,\,\forall i=1,\dots,N$ converge to zero. This completes the proof. \hfill $\square$

\medskip

\noindent
\textbf{Remark 4.} The attacker can deceive the existing $H_{\infty}$  controller by using its IMP-based adversarial input. Although the global performance function  $\Gamma(k)$ reflects the adverse effect of the attacks, the local neighborhood tracking error does not. Therefore, the local performance measure does not ensure  the global performance of the DMAS under the influence of the sophisticated attacks. This analysis reinforces the design of resilient control protocol to mitigate the adverse effects of the attack.

\smallskip

\section{Resilient Distributed Control Protocol for Attacks on Sensor and Actuator : An Adaptive Approach }

This section presents the design of a resilient distributed control protocol for the mitigation of the adverse effect of attacks on sensors and actuators of an agent in the discrete-time DMAS. Regardless of the magnitude of attack $f(k)$ on sensors and actuators of an agent and its reachability from intact agents, our distributed adaptive compensator is resilient against attacks and avoids catastrophic effects. To this end, first, the expected normal behavior of each agent is predicted using an observer-like predictor (called here expected state predictor), which employs the agent's dynamics to predict its expected normal state at each time step. This expected state predictor does not use any actual state measurement, and, instead, calculates the expected normal state of the agent based on the evolution rule of its dynamics, and taking into account the local information it receives from its neighbors. Then, a distributed adaptive compensator is designed using predicted behavior of agents to compensate for any discrepancy between the actual state and its predicted normal one.\par
Consider the  estimated  state  for agent $i$ as  $\hat{x}_i(k)$. The distributed expected state predictor is designed  as
\begin{equation}
\begin{gathered}
  {{\hat x}_i}(k + 1) = A{{\hat x}_i}(k)\,\, + \,cBK(1+ h_i)^{-1}\sum\limits_{j = 1}^N {{a_{ij}}({{\hat x}_j} - {{\hat x}_i})}   \hfill \\
\end{gathered} 
\label{d11a}
\end{equation}
where the gain $K$ and the coupling coefficient $c$ are to be designed to ensure $A_c$ in (\ref{eq_5}) is Schur. The global expected state predictor state vector for (\ref{d11a}) can be written as  $\hat{x}(k)=[\hat{x}_{1}^{T}(k), \hat{x}_{2}^{T}(k),\dots,\hat{x}_{N}^{T}(k)]^T \in R^{nN}$.
\smallskip

\noindent
\textbf{Lemma 3.} Consider the $N$ expected state predictors given in (\ref{d11a}). Let the feedback gain $K$ and coupling coefficient $c$ are designed to ensure $A_c$ in  (\ref{eq_5}) is Schur. Then, the expected state predictor state $\hat{x}(k)$ converges to the desired consensus value.

\smallskip

\noindent
\textit{Proof.}  The designed expected state predictor in  (\ref{d11a}) can be expressed as  
\vspace{-0.3cm}
\begin{equation}
{{\hat x}_i}(k + 1) = A{{\hat x}_i}(k) + B\hat{u}_i(k)\,\,\,\,\,\,
\label{d3}
\end{equation}
where 
\begin{equation}
\hat{u}_i(k) = cK\hat{\varepsilon} _i(k)
\label{d4}
\end{equation}
with the local neighborhood tracking error $\hat{\varepsilon}(k)$ as
\begin{equation}
\hat{\varepsilon}_i(k) =(1+ h_i)^{-1}\sum\limits_{j = 1}^N {{a_{ij}}({{\hat x}_j} - {{\hat x}_i})})
\label{d5}
\end{equation}
One can write the global expected state predictor state dynamics as 
\begin{equation}
   \hat{x}(k+1) = A_c \hat{x}(k) \in R^{nN}
    \label{d6}
\end{equation}
which yields
\begin{equation}
   \hat{x}(k) = A_c^k \hat{x}(0) \in R^{nN}
    \label{d61}
\end{equation}

As $A-c\lambda_{i}BK$  is Schur stable, with $\lambda_i$ be the eigenvalues of the normalized graph Laplacian matrix $\hat{L}$ for $i=2,\dots,N$ and $\lambda_1=0$. Therefore, the expected state predictor states achieve the desired consensus value and written as
\begin{equation}
\hat{x}(k) = ({r^T \otimes A^k})\left[ \begin{matrix}
{\hat{x}_1(0)}  \\ 
   .  \\ 
   .  \\ 
   {\hat{x}_N(0)}  \\ 
\end{matrix} \right]\,\,i=1,\dots,N\,\,\,as\,\,\,k \to \infty
\label{eq_d61}
\end{equation} \hfill $\square$


\textbf{Remark 5.} Note that a broad class of the DMAS includes the leader-follower  or the containment control problem (i.e. MAS with multiple-leader) for which even if the  $\hat{x}_i(0) \neq x_i(0)$, Lemma $3$ is valid. This is because, the reference trajectory to be followed by agents is determined by the leaders, which are assumed to be trusted by using more advanced sensors and investing more security. The system (\ref{d11a}) acts as a reference model for the agents and if $\hat{x}_i(0) \neq x_i(0)$, even for the intact DMAS, $d_i$ in (\ref{d9}) will be nonzero until the difference between the initial conditions is gone. Agents converge to the desired behavior irrespective of initial values. 


\smallskip

Although the attacks on actuators and/or sensors can adversely affect the agents dynamics, they cannot affect the dynamics of the distributed expected state predictor (\ref{d11a}), unless they entirely compromise the agent which is extremely harder to do for the attacker.

The deviation of the agent's behavior from the normal behavior is estimated by distributed expected state predictor. Then, an adaptive attack compensator is developed using an expected state predictor. The designed adaptive compensator is augmented with the controller for the mitigation of the adversarial input.\par

\begin{figure}[H]
\begin{center}
\includegraphics[width=85mm,height=75mm]{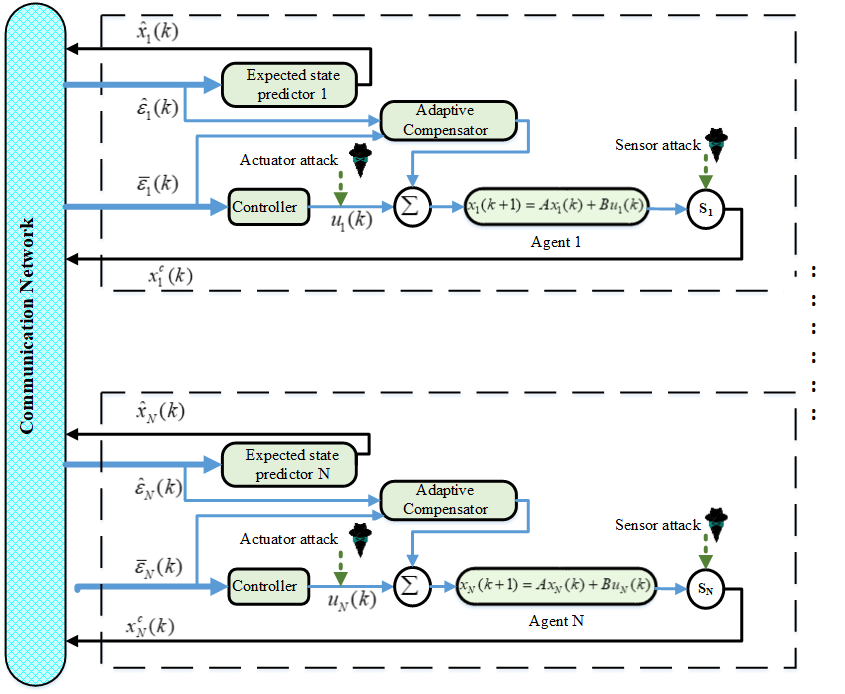}
\end{center}
\caption{Architecture of the proposed adaptive resilient controller. $S_i$ represents the sensor of agent $i \forall i=1,\dots,N.$}
\label{1}
\end{figure}

In contrast to existing detection-removing approaches \cite {c32}-\cite {c36}, which require a strong network connectivity, the developed resilient distributed controller preserves network topology and achieves the desired consensus without any restrictions on the number of agents under sensor and actuator attacks. Attacks on communication links i.e. denial of service (DoS) attack can be mitigated by integrating existing attack detection/identification methodologies \cite {c32}-\cite {c36} with the proposed approach. Therefore, agents under the influence of the adversarial input on sensors and actuators can be recovered using the proposed resilient controller and be brought back to the network in intact mode without being isolated.\par

\smallskip

We now design a distributed resilient control protocol as
\begin{equation}
u_{i,r}(k)=u_{i}(k)+u_{i,comp}(k)
\label{w11}
\end{equation}
where, $u_{i}(k)$ represents standard control protocol defined in (\ref{eq_3}) and $u_{i,comp}(k)$ represents the distributed adaptive compensator protocol responsible for rejection of the adversarial input.\par

\smallskip

Consider the feedback gain $K$  in the control protocol (\ref{eq_3}) given as 
\begin{equation}
K = {({R_1} + {B^T}{P_1}B)^{ - 1}}{B^T}{P_1}A = \bar{R}_1^{-1}{B^T}{P_1}A
\label{d7}
\end{equation}
where $R_1$ is a positive definite design matrix, and $P_1$ is solution of
\begin{equation}
{A^T}{P_1}A - {P_1} - {A^T}{P_1}B{({R_1} + {B^T}{P_1}B)^{ - 1}}{B^T}{P_1}A = {Q_1}
\label{d8}
\end{equation}
with a positive definite matrix $Q_1$.

The designed distributed control protocol is given by 
\begin{equation}
u_{i,r}(k)=cK\bar{\varepsilon}_i(k)-d_i(k)
\label{d9}
\end{equation}
where $d_i(k)$ is the estimated response of the adaptive compensator and $K$ is the gain given by  (\ref{d7}) and (\ref{d8}). The local neighborhood tracking error $\bar{\varepsilon}_i(k)$ in (\ref{d9}) is given by
\begin{equation}
\bar{\varepsilon}_i(k)=(1+ h_i)^{-1}\sum\limits_{j =1}^N {{a_{ij}}({x_j^c}(k) - {x_i^c}(k))}
\label{d91a}
\end{equation}
The update for the distributed adaptive compensator is designed as
\begin{equation}
d_i(k+1)=\theta cK(\hat {\varepsilon}_i(k) -\bar{\varepsilon}_i(k))+\theta d_i(k)
\label{d10}
\end{equation}
where $\theta>0$ is a design parameter, and $\bar{\varepsilon}_i(k)$ and  $\hat{\varepsilon}_i(k)$ are defined in (\ref{d91a}) and (\ref{d5}). 


\smallskip

\noindent
\textbf{Remark 6.} The information exchanged among agents in (\ref{d91a}) is the corrupted state measurement $x_i^c(k)$ defined in (\ref{eq_8}) which is different from $x_i(k)$ defined in (\ref{eq_9}). However, if the agent $i$ is intact or only under actuator attack, then  $x_i^c(k)=x_i(k)$. If attacker corrupt the sensor data, then $x_i^c(k) \neq x_i(k)$. In most of the existing DMAS work, it is assumed the states of the agents are measurable and also we consider the same in our work. Therefore, the attack on the sensor corrupts the state measurements.  



According to Lemma 3, the expected state predictor converges to the desired consensus value. Therefore, consensus of DMAS can be achieved by showing the convergence of the agent state $x_i(k)$  to the predicted state $\hat{x}_i(k)$. Define the consensus error $\tilde{x}(k)$ as
\vspace{-0.15cm}
\begin{equation}
\tilde{x}(k)=x(k)-\hat{x}(k)
\label{d12}
\end{equation}

In the following theorem, we show that the consensus error remains bounded using the proposed resilient adaptive controller.\par

\smallskip

\noindent
\textbf{Theorem 5.}  {Consider the DMAS (\ref{eq_9}) under attacks on sensors and actuators. Let the control protocol be developed as (\ref{d9})-(\ref{d10}). Then, the agent's consensus errors defined in (\ref{d12}) are bounded, and the bound can be made arbitrarily small, despite the attack.} 

\smallskip 
 
\noindent
\textit{Proof.} According to Lemma 4, the expected state predictor converges to the desired consensus value. Therefore, consensus of discrete-time DMAS can be achieved by showing the convergence of the agent state $x_i(k)$  to the predicted state $\hat{x}_i(k)$. 
Then, with (\ref{eq_9}) and (\ref{d3}), one can write $\tilde{x}(k+1)$ as
\begin{equation}
\tilde x(k + 1) = ({I_N} \otimes A - c\hat L \otimes BK)\tilde x(k) - ({I_N} \otimes B)\tilde d(k)
\label{d13}
\end{equation}
where 
\vspace{-0.2cm}
\begin{equation}
\tilde{d}(k)=d(k)-f(k)
\label{d14}
\end{equation}
denotes attack rejection error with  $d(k)=[d_{1}^{T}(k), d_{2}^{T}(k),\dots,d_{N}^{T}(k)]^T \in R^{nN}$ as the global adaptive compensator vector and the dynamics of the attack $f(k)$ is defined in (\ref{eq_20}).\par
Using (\ref{d10}), the global dynamics of the adaptive compensator can be written as
\begin{equation}
d(k + 1) = \theta c\hat{L}\otimes {{\bar{R}_1}^{-1}}{B^T}{P_1}A\tilde x(k) + \theta \tilde{d}(k) + \theta \bar{f}(k)
\label{d15}
\end{equation}
where $\bar{R}_1=R_1+B^TP_1B$ and $\bar{f}(k)=2f(k)-(\gamma \otimes I_N)u^{a}$. Note that  $\bar{f}(k)=f(k)$ only if the actuator of the agent is compromised. Define $Q_2=Q_{2}^T>0$  as $Q_2=cR_2(I+H)^{-1}L=cR_2\hat{L}$
with some positive definite $R_2$. Let the real part of the minimum eigenvalue of the normalized graph Laplacian matrix $\hat{L}$ be $\lambda_{m}$.\par
Define the Lyapunov candidate function function as 
\begin{equation}
V(k) = {\tilde x^T}(k)({Q_2} \otimes {P_1})\tilde x(k) + {\theta ^{ - 2}}{\tilde d^T}(k)({R_2} \otimes {\bar R_1})\tilde d(k)
\label{d16}
\end{equation}
The difference equation of the Lyapunov candidate function can be written as 
\begin{align}
\begin{gathered}
\Delta V(k) = V(k + 1) - V(k) \\
= \underbrace {{{\tilde x}^T}(k + 1)({Q_2} \otimes {P_1})\tilde x(k + 1) - {{\tilde x}^T}(k)({Q_2} \otimes {P_1})\tilde x(k)}_{part\,\,1} \\
+ \underbrace {{\theta ^{ - 1}}{{\tilde d}^T}(k + 1)({R_2} \otimes {R_1})\tilde d(k + 1) - {\theta ^{ - 1}}{{\tilde d}^T}(k)({R_2} \otimes {R_1})\tilde d(k)}_{part\,\,2}
\end{gathered}
\label{d17}
\end{align}
Using (\ref{d13}), part $1$ of the difference equation of the Lyapunov candidate function  (\ref{d17}) can be expressed as
\begin{align}
\begin{gathered}
= {{\tilde x}^T}(k)({Q_2} \otimes {A^T}{P_1}A  - 2c{Q_2}\hat L \otimes {A^T}{P_1}BK \hfill \\  + {c^2}{{\hat L}^T}{Q_2}\hat L \otimes {(BK)^T}{P_1}BK 
- ({Q_2} \otimes {P_1}))\tilde x(k) \hfill \\ - 2{{\tilde x}^T}(k)[{Q_2} \otimes {A^T}{P_1}B  - c{{\hat L}^T}{Q_2} \otimes {(BK)^T}{P_1}B]\tilde d(k) \hfill \\
+ {{\tilde d}^T}(k)({Q_2} \otimes {B^T}{P_1}B)\tilde d(k) \hfill \\ 
\end{gathered}
\label{d18}
\end{align}
Using the Young's inequality, one can further simplify and express (\ref{d18}) as 
\begin{align}
\begin{gathered}
\leqslant - {{\tilde x}^T}(k)({Q_2} \otimes {Q_1})\tilde x(k) - {{\tilde x}^T}(k)( - {Q_2} + 2c{Q_2}\hat L) \otimes {A^T}{P_1}BK)\tilde x(k) \hfill \\
  \,\,\,\,\,\,\,\, + 2{c^2}{\lambda _{\min }}({c^2}{{\hat L}^T}\hat L{\lambda _{\min }}(TQ_1^{ - 1})){{\tilde x}^T}(k)({Q_2} \otimes {Q_1})\tilde x(k) \hfill \\
  \,\,\,\,\,\,\,\, - 2{{\tilde x}^T}(k)({Q_2} \otimes {A^T}{P_1}B)\tilde d(k) + 2{{\tilde d}^T}(k)({Q_2} \otimes {B^T}{P_1}B)\tilde d(k) \hfill \\ 
\end{gathered} 
\label{d19}
\end{align}
where $T=K^TB^TP_1BK$. We now consider the  part $2$ of the difference equation of the Lyapunov candidate function  in (\ref{d17}) as
\begin{align}
{\theta ^{ - 2}}{{\tilde d}^T}(k + 1)({R_2} \otimes {{\bar R}_1})\tilde d(k + 1) - {\theta ^{ - 2}}{{\tilde d}^T}(k)({R_2} \otimes {{\bar R}_1})\tilde d(k)
\label{d20}
\end{align}
where ${{\bar R}_1} = ({R_1} + {B^T}{P_1}B)$ is a positive definite matrix. Using (\ref{d14}), one can express (\ref{d20}) as
\begin{align}
\begin{gathered}
\frac{1}{\theta^2}[{d^T}(k + 1)({R_2} \otimes {{\bar R}_1})d(k + 1) - 2{d^T}(k + 1)({R_2} \otimes {{\bar R}_1})f(k + 1) \hfill \\ + {f^T}(k + 1)({R_2} \otimes {{\bar R}_1})(f(k + 1) - {{\tilde d}^T}(k)({R_2} \otimes {{\bar R}_1})\tilde d(k)]
\end{gathered}
\label{d21}
\end{align}
Using the dynamics of the distributed adaptive compensator in (\ref{d15}) with (\ref{d21}), one has
\begin{align}
\begin{gathered}
= {{\tilde x}^T}(k)(c{{\hat L}^T}{Q_2} \otimes {K^T}{B^T}{P_1}A)\tilde x(k) + 2{{\tilde d}^T}(k)({Q_2} \otimes {B^T}{P_1}A)\tilde x(k) \hfill \\ +2[{\bar{f}}(k)-{\theta ^{ - 1}}{f}(k + 1)]^T({Q_2} \otimes {B^T}{P_1}A)\tilde x(k)\hfill \\+ (1 - {\theta ^{ - 2}}){{\tilde d}^T}(k)({R_2} \otimes {{\bar R}_1})\tilde d(k) \hfill \\+ [{\bar{f}}(k)-{\theta ^{ - 1}}{f}(k + 1)]^T({R_2} \otimes {{\bar R}_1})\tilde d(k)\hfill \\+ [{\bar{f}}(k)-{\theta ^{ - 1}}{f}(k + 1)]^T({R_2} \otimes {{\bar R}_1})[{\bar{f}}(k)-{\theta ^{ - 1}}{f}(k + 1)]\hfill \\
\end{gathered} 
\label{d23}
\end{align}
Using the Young's inequality, one can simplify (\ref{d23}) as 
\begin{align}
\begin{gathered}
\leq \frac{3}{2}{{\tilde x}^T}(k)(c{Q_2}\hat L \otimes {A^T}{P_1}BK)\tilde x(k) + 2{{\tilde d}^T}(k)({Q_2} \otimes {B^T}{P_1}A)\tilde x(k) \hfill \\
  + (2 - {\theta ^{ - 2}}){{\tilde d}^T}(k)({R_2} \otimes {{\bar R}_1})\tilde d(k) \hfill \\+ 4[{\bar{f}}(k)-{\theta ^{ - 1}}\psi(k){f}(k)]^T({R_2} \otimes {{\bar R}_1})[{\bar{f}}(k)-{\theta ^{ - 1}}\psi(k){f}(k)]\hfill \\ 
\end{gathered} 
\label{d24}
\end{align}
where $\psi(k)$ denotes how the value of attack signal changes at next time instant. If attack signal is constant, i.e., $f(k+1)=f(k)$, then $\psi(k)=1$. Thus, one can infer that $\psi(k)$ is always bounded, i.e., $|\psi(k)|<\zeta \,\,\, \forall \,\,k$. Integrating equation (\ref{d19}) and (\ref{d24}) with further simplification, one has 
\begin{align}
\begin{gathered}
\Delta V \leqslant  - {{\tilde x}^T}(k)({Q_2} \otimes {Q_1})\tilde x(k) \hfill \\ - {{\tilde x}^T}(k)( - {Q_2} + \frac{1}{2}c{Q_2}\hat L) \otimes {A^T}{P_1}BK)\tilde x(k) \hfill \\  + 2{c^2}{\lambda _{\min }}({{\hat L}^T}\hat L){\lambda _{\min }}(TQ_1^{ - 1})){{\tilde x}^T}(k)({Q_2} \otimes {Q_1})\tilde x(k) \hfill \\
- \,({\theta ^{ - 2}} - 2 - 2{\lambda _{\min }}(c\hat{L}{B^T}{P_1}B\bar R_1^{ - 1}){{\tilde d}^T}(k)({R_2} \otimes {{\bar R}_1})\tilde d(k) \hfill \\  + 4[{\bar{f}}(k)-{\theta ^{ - 1}}\zeta{f}(k)]^T({R_2} \otimes {{\bar R}_1})[{\bar{f}}(k)-{\theta ^{ - 1}}\zeta{f}(k)]\hfill \\ 
\end{gathered} 
\label{d26}
\end{align}
One can show that $\Delta V \leq 0$, if the coupling coefficient satisfies $\frac{2}{{{\lambda _m}}} < c < \frac{1}{{{\lambda _m}\sqrt {2{\lambda _{\min }}(TQ_1^{ - 1})} }}$ and 
\begin{align}
\left\| {\tilde d(k)} \right\|> \frac{{4\left\| {(\bar f(k) - {\theta ^{ - 1}}\zeta{f}(k))} \right\|}}{{{\theta ^{ - 2}} - 2 - 2{\lambda _{\min }}(c\hat L{B^T}{P_1}B\bar R_1^{ - 1})}}
\label{d12bb}
\end{align} 
The design parameter $\theta$ can be chosen such $\theta< \frac{1}{{\sqrt {2+{\lambda _{\min }}(c\hat L{B^T}{P_1}B\bar R_1^{ - 1})} }}$ and then, one can ensure the bound in (\ref{d12bb}). This shows that the agent's consensus error is bounded.  Therefore, the actual agent's state $x(k)$ achieve the desired consensus behavior with a bounded error that can be made arbitrarily small by appropriate selection of design parameter $\theta$. This completes the proof. $\hfill$ $\square$


\smallskip

\noindent
{\textbf{Remark 7.} The  coupling coefficient $c$ needs to be in a certain range which depends on the $\lambda_m$ and $\lambda_{min}(TQ_1^{-1})$. This condition is standard in the literature of DMAS \cite{c22}. On the other hand, the condition for the bound on $\tilde{d}(k)$ in (\ref{d12bb}) depends on the design parameters $\theta$, and one can select this parameter to satisfy (\ref{d12bb}) which ensures $\Delta V \leq 0$. Thus, the bound on consensus error can be made arbitrarily small based on selection of design parameter $\theta$. Moreover, this bound is conservative, and as shown in the simulation results, the consensus error almost goes to zero.}

\smallskip

\noindent
\textbf{Remark 8.} As presented in Theorem $4$ and Corollary $2$,  existing  $H_{\infty}$ approaches minimize the local neighborhood tracking error of the system $\epsilon_i(k)$ and are not capable of attenuating sophisticated attacks. In contrast, the designed distributed resilient control can successfully attenuate the adverse effects of attacks using the distributed adaptive compensator. The developed compensator  $d_i(k)$ in  (\ref{d15}) minimizes  the deviation of the local neighborhood tracking error of the system $\epsilon_i(k)$ from the local neighborhood tracking error of the expected state predictor $\hat{\epsilon}_i(k)$. We can also infer that, although the proposed controller is designed for leaderless multi-agent systems, it can be used for the leader-follower systems and the containment control systems.   

\smallskip

\noindent
\textbf{Remark 9.} Compromised agents under the effect of the sensor attack might not be recovered completely and result a non-zero bound in (\ref{d26}). The proposed distributed adaptive law compensates the difference between the incoming neighboring sensor measurement $(x_{i}^c(k))$ and the desired state $\hat{x}_i(k)$ and   $x_{i}^c(k) \neq {x}_i(k)$ in the case of sensor attack. Under the actuator attack $x_{i}^c(k) = {x}_i(k)$ and the error bound for (\ref{d26}) can be made arbitrarily small.

\vspace{-0.2cm}

\section{Simulation Results}
This section presents simulation results to validate the effectiveness of the presented work. We consider both leaderless as well as leader-follower network for the simulation. First, a leader-follower network of autonomous underwater vehicle's (AUV's) in  Fig. \ref{f0} is considered for the evaluation of the presented results. Then, the leaderless network is considered.

\vspace{-0.3cm}

\subsection{Leader-follower Network}
\begin{figure}[H]
\begin{center}
\includegraphics[width=82mm,height=55mm]{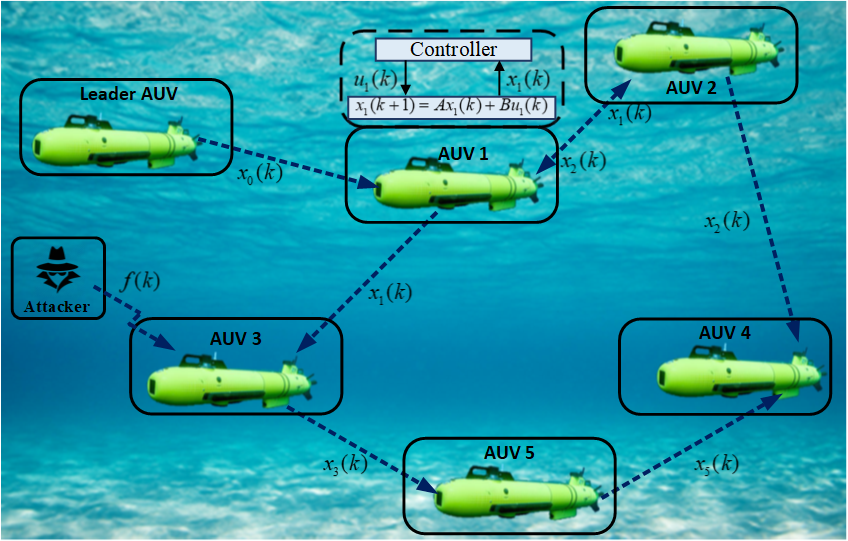}
\end{center}
\caption{Distributed network of AUVs under the influence of attack}
\label{f0}
\end{figure}

The communication network is shown in Fig.\ref{f0} for a set of Sentry AUVs. Sentry AUV is manufactured by the Woods Hole Oceanographic Institution \cite{cauv}. The linearized model of the Sentry is of 6 DOF, but it is generally decomposed into four non-interacting subsystems which are speed subsystem $(u)$,  the roll subsystem $(\phi)$, the steering subsystem $(\nu, r,\psi)$, the diving subsystem $(\omega, q, z, \theta)$. Here, we focus on the diving subsystem of Sentry AUV for the desired depth maneuvering in the leader-follower network. 

\vspace{-0.3cm}

\begin{figure}[H]
\begin{subfigure}[b]{\columnwidth}
\includegraphics[width=0.72\columnwidth]{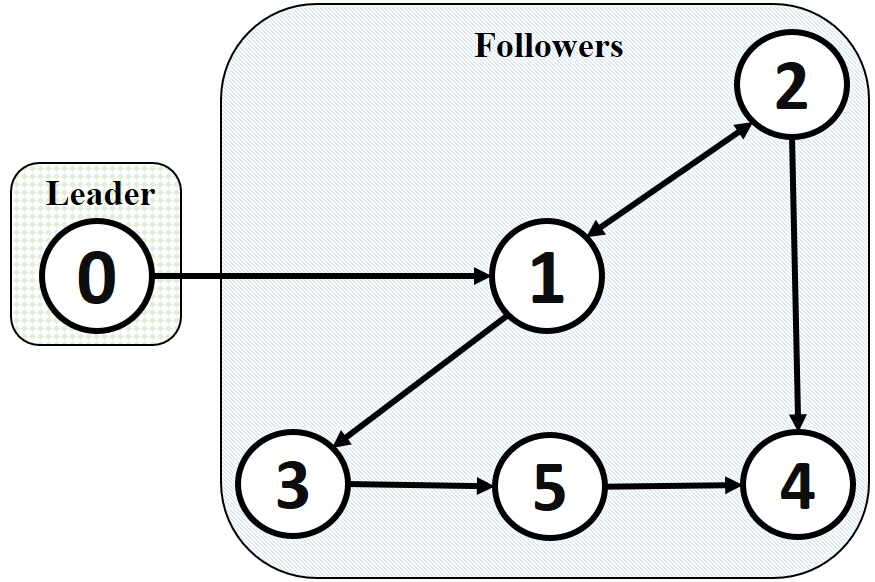} 
\end{subfigure}
\caption{Graph topology}
\label{f1}
\end{figure}

\noindent
The graph topology for Fig. \ref{f0} is shown in Fig. \ref{f1} for the team of Sentry AUVs communicating with each other with the following diving subsystem dynamics
\vspace{-0.2cm}
\[{x_i}(k + 1) = A{x_i}(k) + B{u_i}(k)\]
where 
\vspace{-0.3cm}
\[{\text{A  =  }}\left[ {\begin{array}{*{20}{c}}
  {{\text{ 0}}{\text{.65}}}&{{\text{0}}{\text{.54}}}&{\text{0}{\text{.0}}}&{{\text{-0}}{\text{.0019}}} \\ 
  {{\text{0}}{\text{.21}}}&{{\text{1}}{\text{.48}}}&{\text{0}{\text{.0}}}&{{\text{-0}}{\text{.01}}} \\ 
   {{\text{ 0}}{\text{.83}}}&{{\text{ 0}}{\text{.84}}}&{{\text{1}}{\text{.0}}}& {{\text{0}}{\text{.99}}}\\ 
   {{\text{0}}{\text{.11}}}& {{\text{1}}{\text{.21}}}& {{\text{0}}{\text{.0}}}& {{\text{0}}{\text{.99}}} 
\end{array}} \right]\]and  
\vspace{-0.4cm}
\[B = \left[ {\begin{array}{*{20}{c}}
  {{\text{0}}{\text{.08}}}&{{\text{0}}{\text{.13}}} \\ 
  {{\text{ - 0}}{\text{.13}}}&{{\text{0}}{\text{.20}}} \\ 
  {{\text{ 0}}{\text{.02}}}&{{\text{0}}{\text{.09}}}\\ 
  {{\text{ - 0}}{\text{.07}}}&{{\text{0}}{\text{.09}}}
\end{array}} \right]\]   
with $x_i(k)=[(\omega_i(k), \,\,\,q_i(k), \,\,\,z_i(k), \,\,\,\theta_i(k))]^T$, and ${u_i}(k) = {[\delta _i^b(k),\,\,\,\delta _i^b(k)]^T}$, where $(\omega_i(k), \,\,\,q_i(k), \,\,\,z_i(k), \,\,\,\theta_i(k))$ and $\delta _i^b(k),\,\,\,\delta _i^b(k)$ represent the heave speed, pitch rate, depth and pitch, and bone and stern plane deflections, respectively.

In the network communication graph, we assumed that the agent $0$ represents a active  non-autonomous leader which aim to follow a desired sinusoidal depth trajectory and agents $1$ to $5$ designate the followers. The leader has the control input $u_0(k)=K_0x_0(k)+r(k)$, where $K_0$ is  state feedback gain, $x_0$ denotes the leader state and $r(k)$ represents the desired sinusoidal trajectory, respectively. Since the leader input is non-zero, slightly different discrete-time control protocol from that the one proposed in the paper is used for which the leader exchanges its input signal $u_0$ with its neighbors and agents reach consensus by exchanging states and leader's input. This, however, does not change our attack analysis and mitigation. The state feedback gain $K_0$ is given by 
\vspace{-0.1cm}
\[{K_0} = \left[ {\begin{array}{*{20}{c}}
  {{\text{  - 0}}{\text{.18}}}&{{\text{-2}}{\text{.25 }}}&{{\text{0}}{\text{.13 }}}&{{\text{-0}}{\text{.21}}} \\ 
  {{\text{1}}{\text{.56}}}&{{\text{5}}{\text{.39}}}&{{\text{0}}{\text{.49 }}}&{{\text{1}}{\text{.59}}} 
\end{array}} \right]\]

\begin{figure}[H]
\begin{subfigure}[b]{\columnwidth}
\includegraphics[width=1\columnwidth,height=36mm]{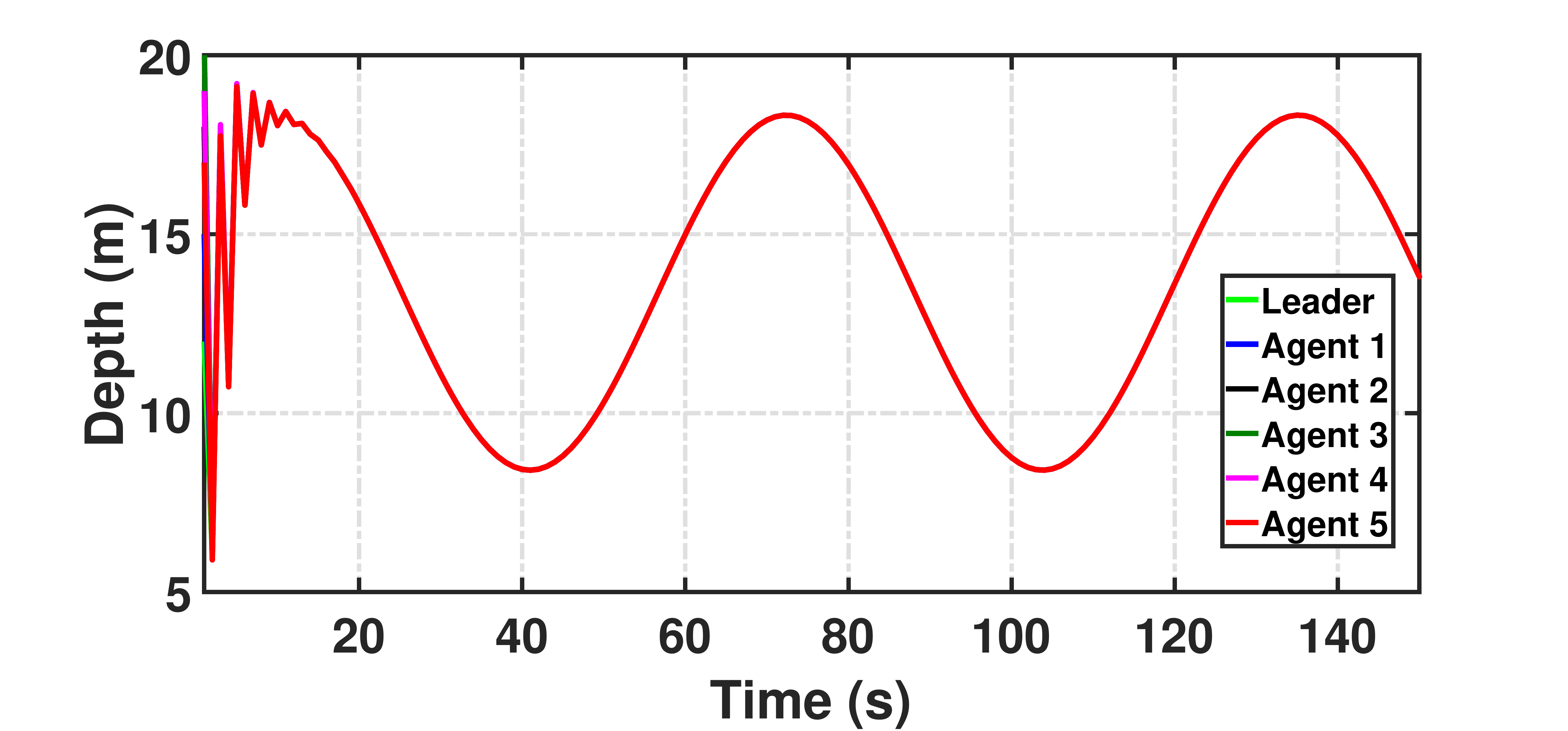} 
\end{subfigure}
\caption{Agents depth trajectory in healthy mode}
\label{lf1}
\vspace{-0.8cm}
\end{figure}

In the absence of the attack, agents follow the desired depth trajectory and illustrate the healthy behavior of the network as shown in Fig. \ref{lf1}. 

\vspace{-0.2cm}

\vspace{-0.5cm}

\begin{figure}[H]
\begin{subfigure}[b]{\columnwidth}
\includegraphics[width=1\columnwidth,height=36mm]{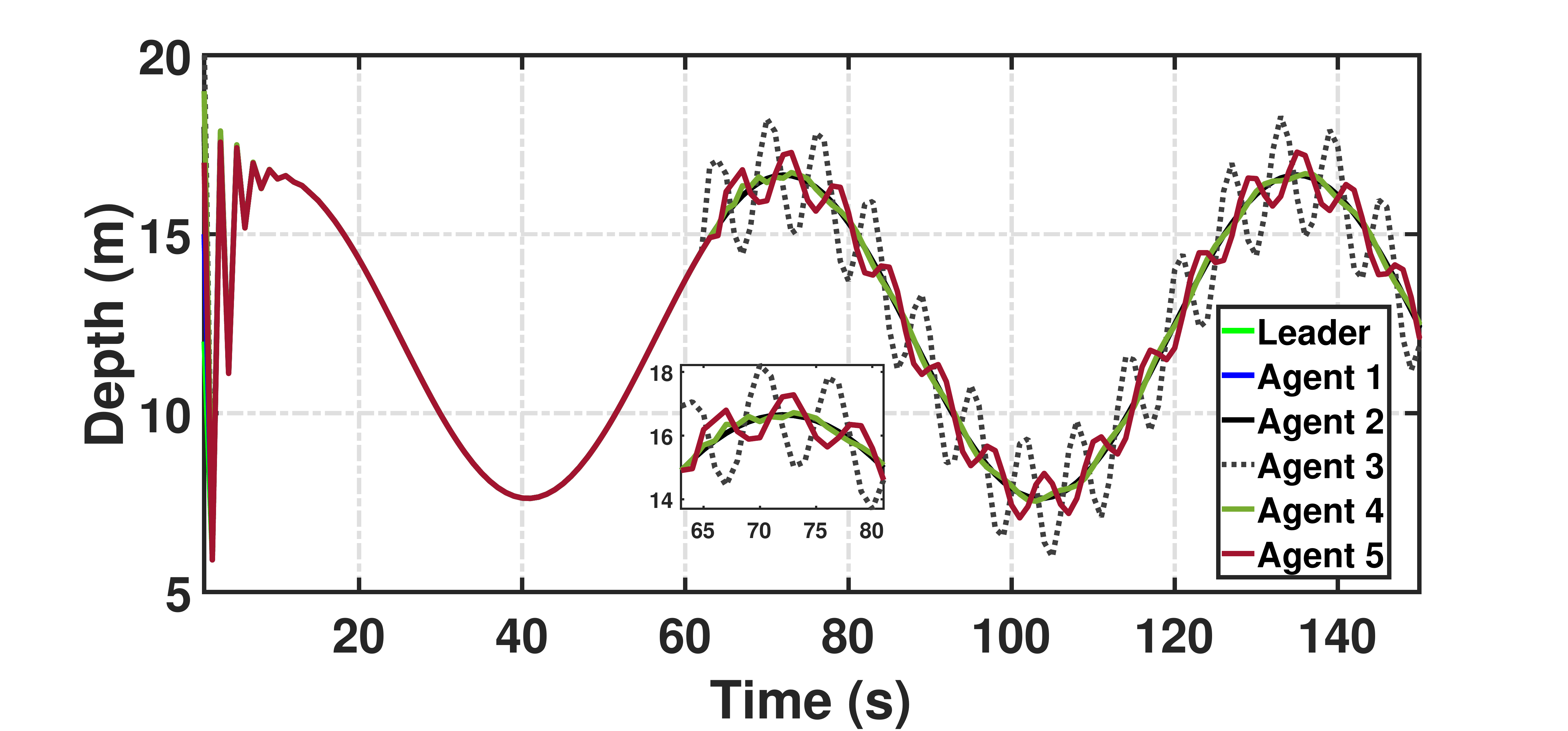} 
\caption{}
\end{subfigure}
\begin{subfigure}[b]{\columnwidth}
\includegraphics[width=1\columnwidth,height=36mm]{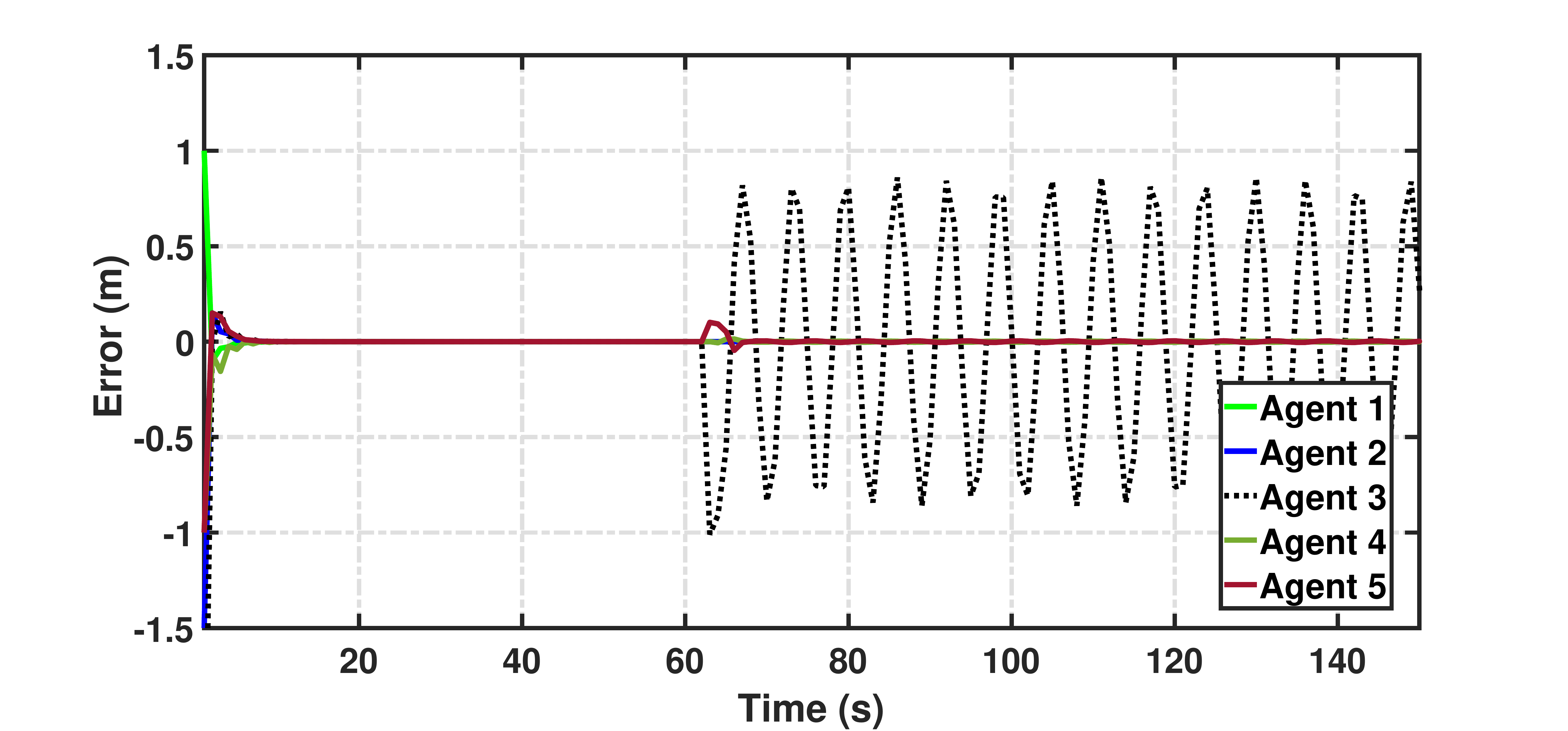}
\caption{}
\end{subfigure} 
\caption{The Agents depth trajectory under the influence of the attack on AUV 3. (a) Depth without adaptive compensator  (b) The local neighborhood tracking error without adaptive compensator}
\label{lf2}
\vspace{-0.4cm}
\end{figure}
Now, the effect of the attack on a non-root node is analyzed. First, we consider the attack on actuators of Agent

\begin{figure}[H]
\begin{subfigure}[b]{\columnwidth}
\includegraphics[width=1\columnwidth,height=45mm]{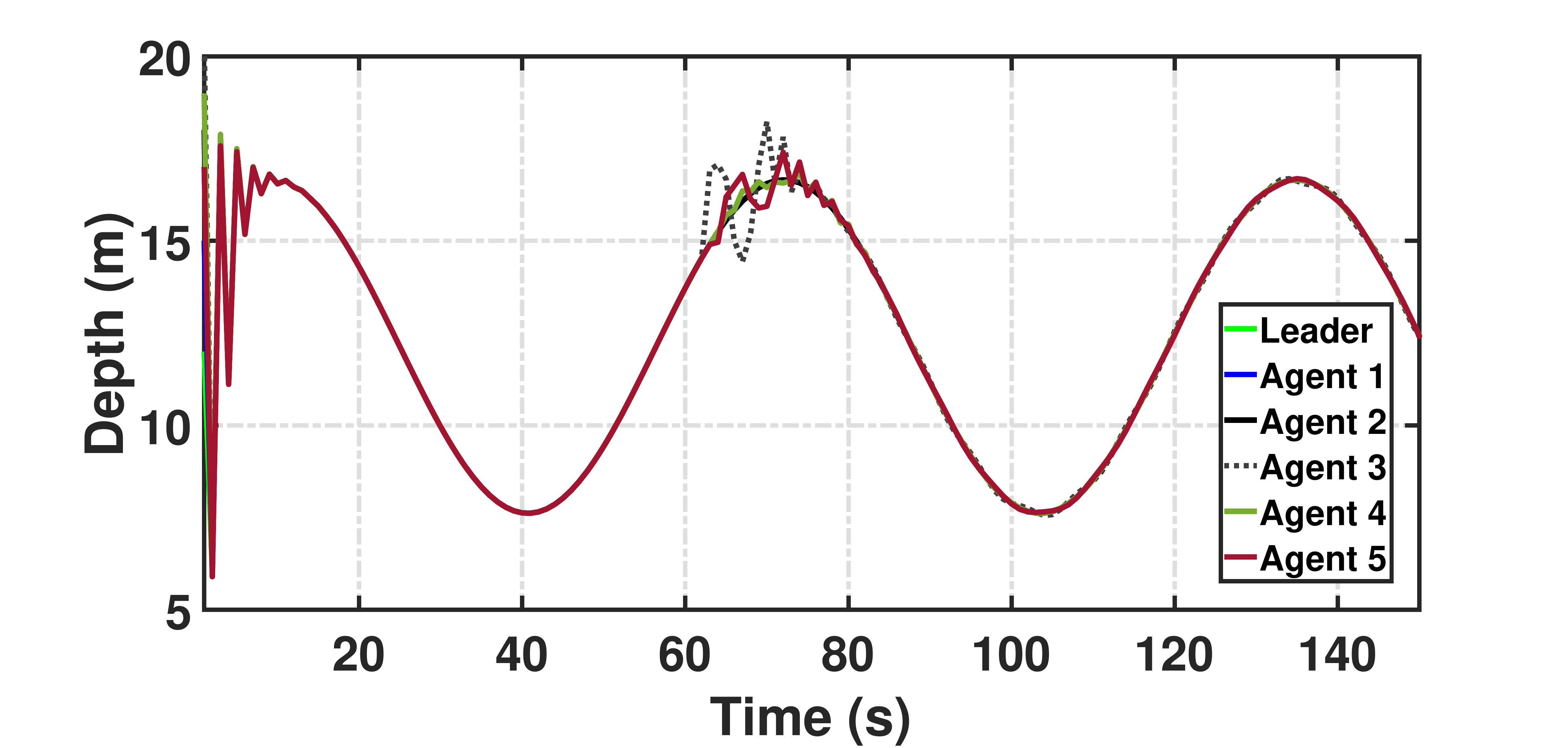} 
\caption{}
\end{subfigure}
\begin{subfigure}[b]{\columnwidth}
\includegraphics[width=1\columnwidth,height=45mm]{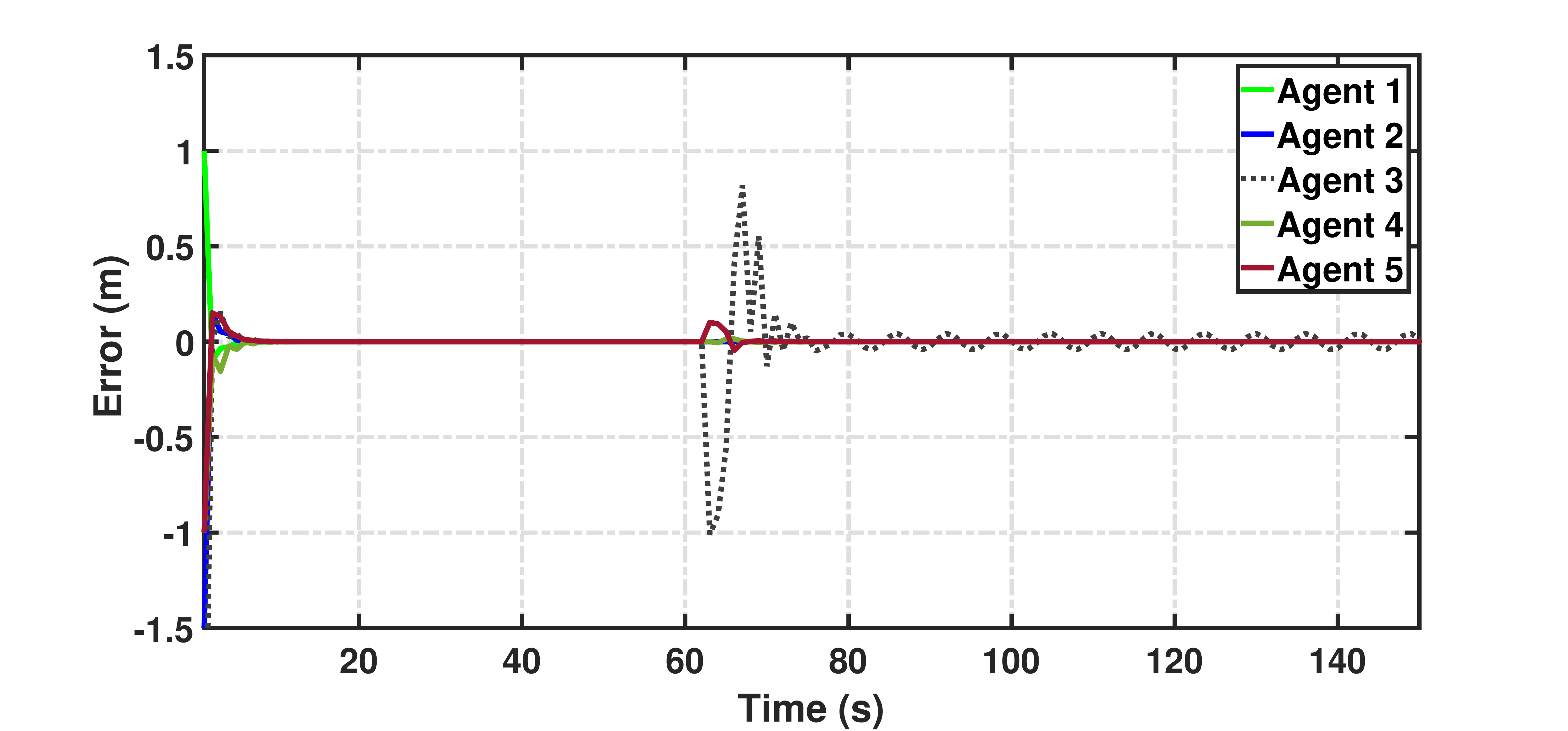}
\caption{}
\end{subfigure} 
\caption{The Agents depth trajectory under the influence of the attack on AUV 3. (a) Depth with adaptive compensator  (b) The local neighborhood tracking error with adaptive compensator}
\label{lf4}
\vspace{-0.3cm}
\end{figure}

\begin{figure}[H]
\begin{subfigure}[b]{\columnwidth}
\includegraphics[width=1\columnwidth,height=45mm]{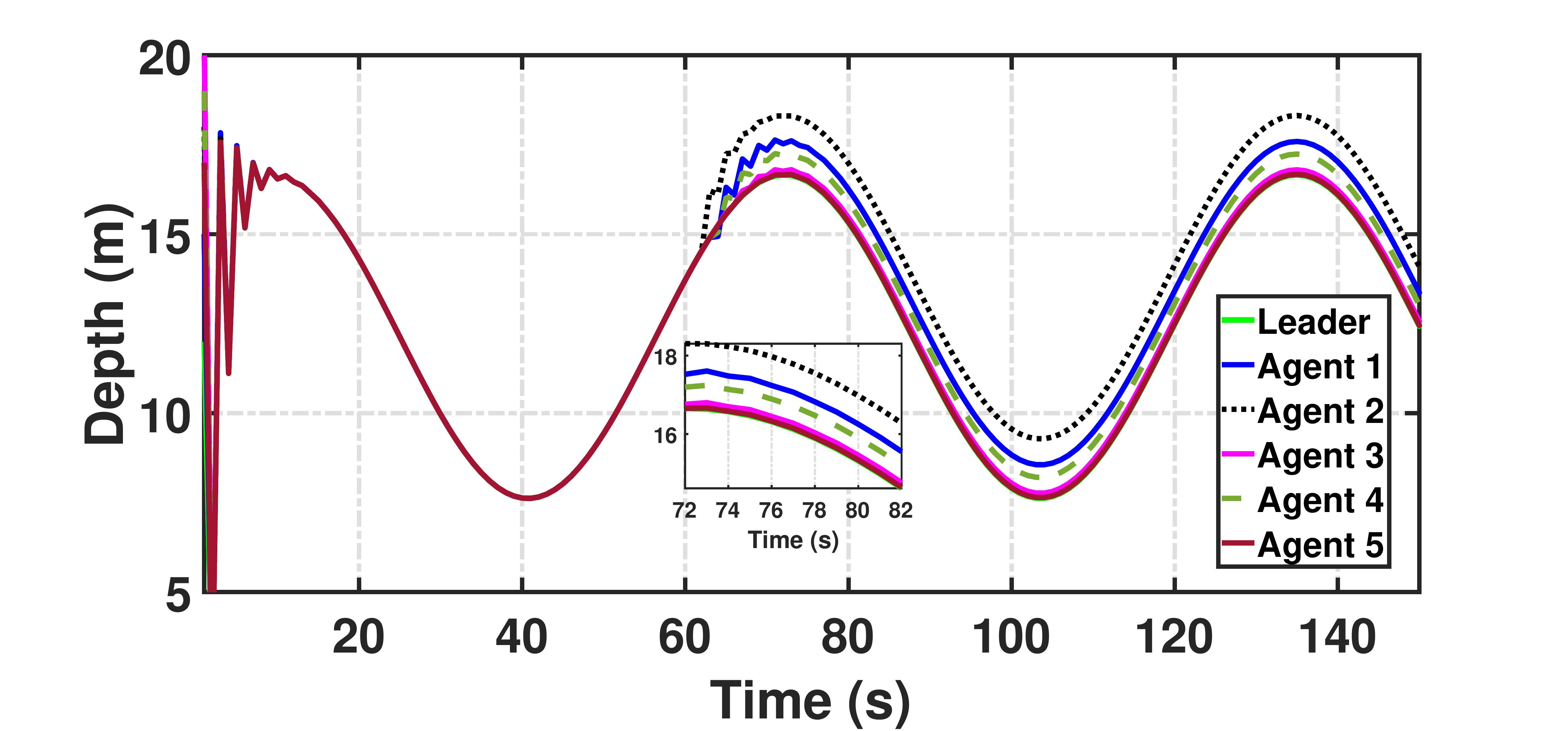} 
\caption{}
\end{subfigure}
\begin{subfigure}[b]{\columnwidth}
\includegraphics[width=1\columnwidth,height=45mm]{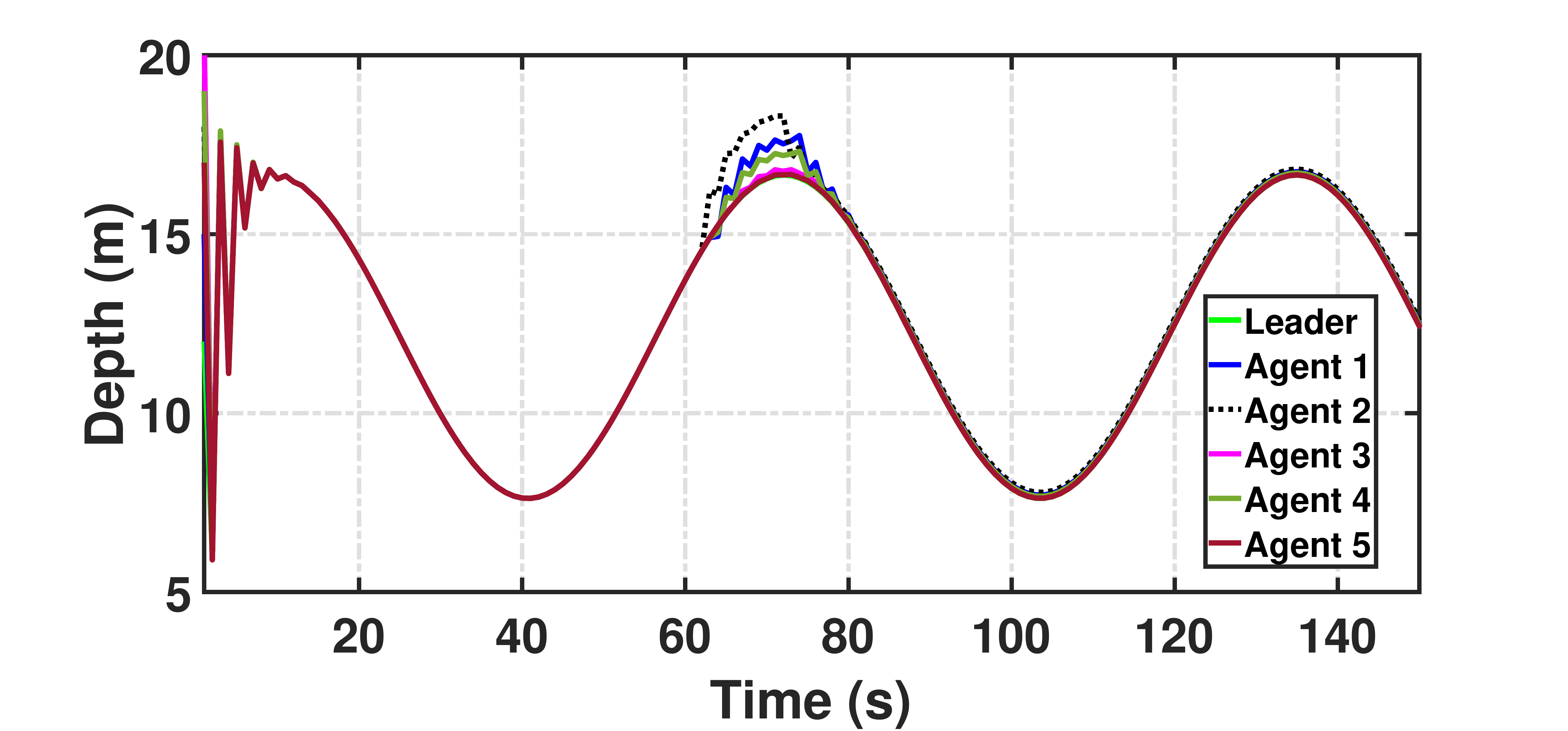}
\caption{}
\end{subfigure} 
\caption{The Agents depth trajectory under the influence of the attack on AUV 2. (a) Depth without adaptive compensator  (b) The local neighborhood tracking error without adaptive compensator}
\label{lf6}
\vspace{-0.3cm}
\end{figure}

\noindent
$3$ (non-root node) with time-varying attack signal $u_{3}^{a}(k) =[10sin(k) \,\,\,\, 10sin(k)]'$ at $t=61$ sec. Fig. \ref{lf2}(a) and (b) show that agents which are reachable from the compromised Agent $3$ are deviated from the desired behavior, despite the local neighborhood tracking error goes to zero for intact agents. These results follow the Theorem $2$ and Theorem $4$.\par

Then, Fig. \ref{lf4}(a) and (b) illustrate the response of the system under the influence of the attacks using the proposed controller (\ref{d9}) from t=$61$ sec with $Q_1$ and $R_1$  be identity matrix in (\ref{d7}) and (\ref{d8}), respectively. The system states achieve the desired consensus behavior and the local neighborhood tracking error goes to almost zero, even in the presence of the attack. These results demonstrate the effectiveness of the proposed resilient controller.  

Now, consider the effect of constant attack signal on actuators of the Agent $2$ (root-node) given by  $u_{2}^{a}(k) =[5\,\,\,\,5]'$ at $t=61$ sec.  Fig. \ref{lf6}(a) shows that the compromised agent affect the reachable agents after attack.  Now, the proposed resilient control protocol in  (\ref{d9}) with adaptive attack compensator in  (\ref{d10}) is applied at t=$61$sec in the distributed network and Fig. \ref{lf6}(b) shows that the system states achieve the desired consensus behavior, even in the presence of the attack. 

\vspace{-0.5cm}

\subsection{Leaderless Network}
Now, consider the same graph topology in Fig. \ref{f1} for leaderless DMAS for $5$ agents without leader Agent $0$. For leaderless system the dynamics of  agent $i$ is considered as 
\begin{align}
\begin{gathered}
  {x_i}(k + 1) = \left[ {\begin{array}{*{20}{c}}
  0&{ - 1} \\ 
  1&0 
\end{array}} \right]{x_i}(k) + \left[ {\begin{array}{*{20}{c}}
  0 \\ 
  1 
\end{array}} \right]{u_i}(k) \hfill \\
  for\,\,\,\,\,\,i = 1,\dots,5 \hfill \\ 
\end{gathered}
\label{d1}
\end{align}
\vspace{-0.2cm}

First,  the effect of the attack on a root node is analyzed with  the IMP-based attack signal. \par

\vspace{-0.5cm}

\begin{figure}[H]
\begin{subfigure}[b]{\columnwidth}
\includegraphics[width=\columnwidth,height=45mm]{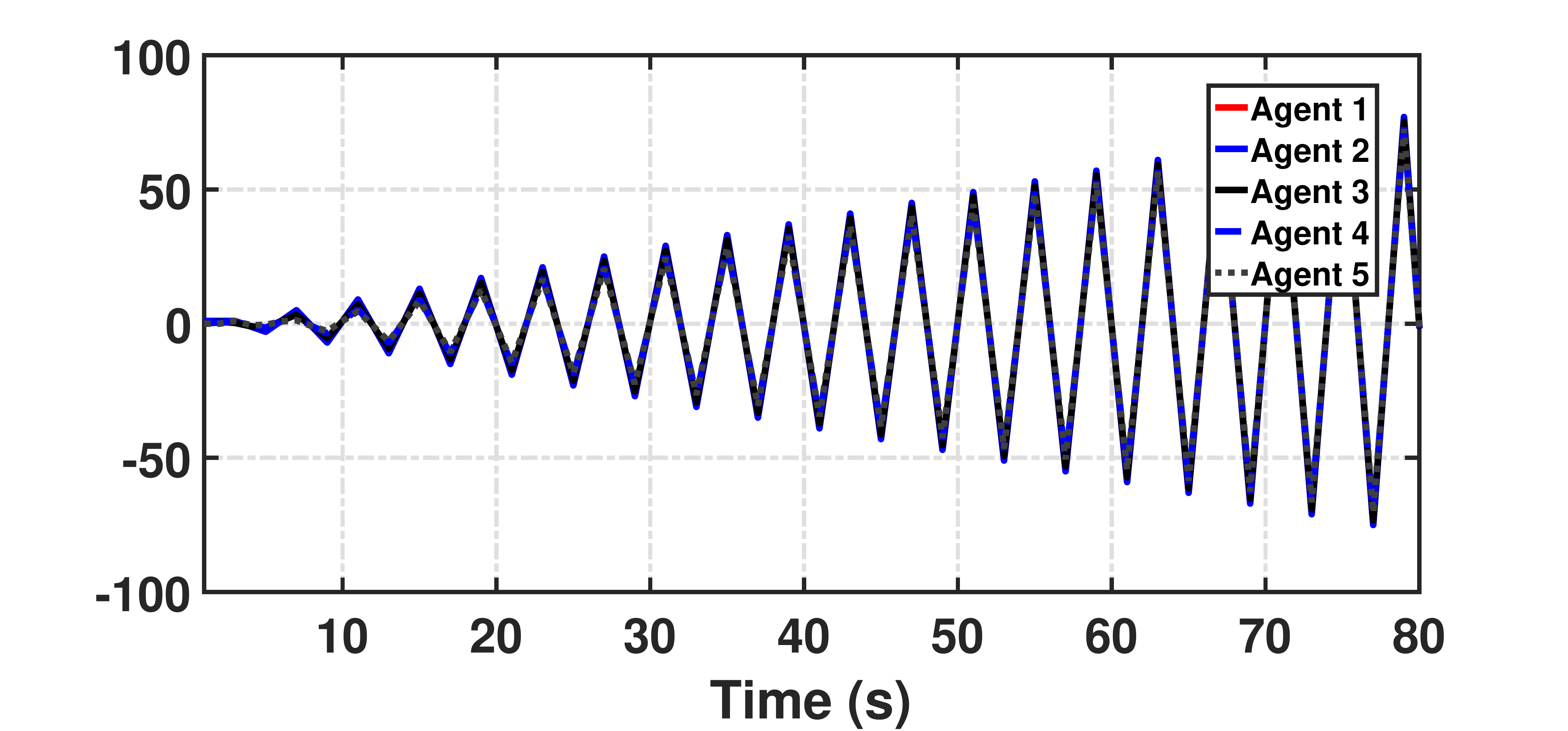} 
\caption{}
\end{subfigure}
\begin{subfigure}[b]{\columnwidth}
\includegraphics[width=\columnwidth,height=45mm]{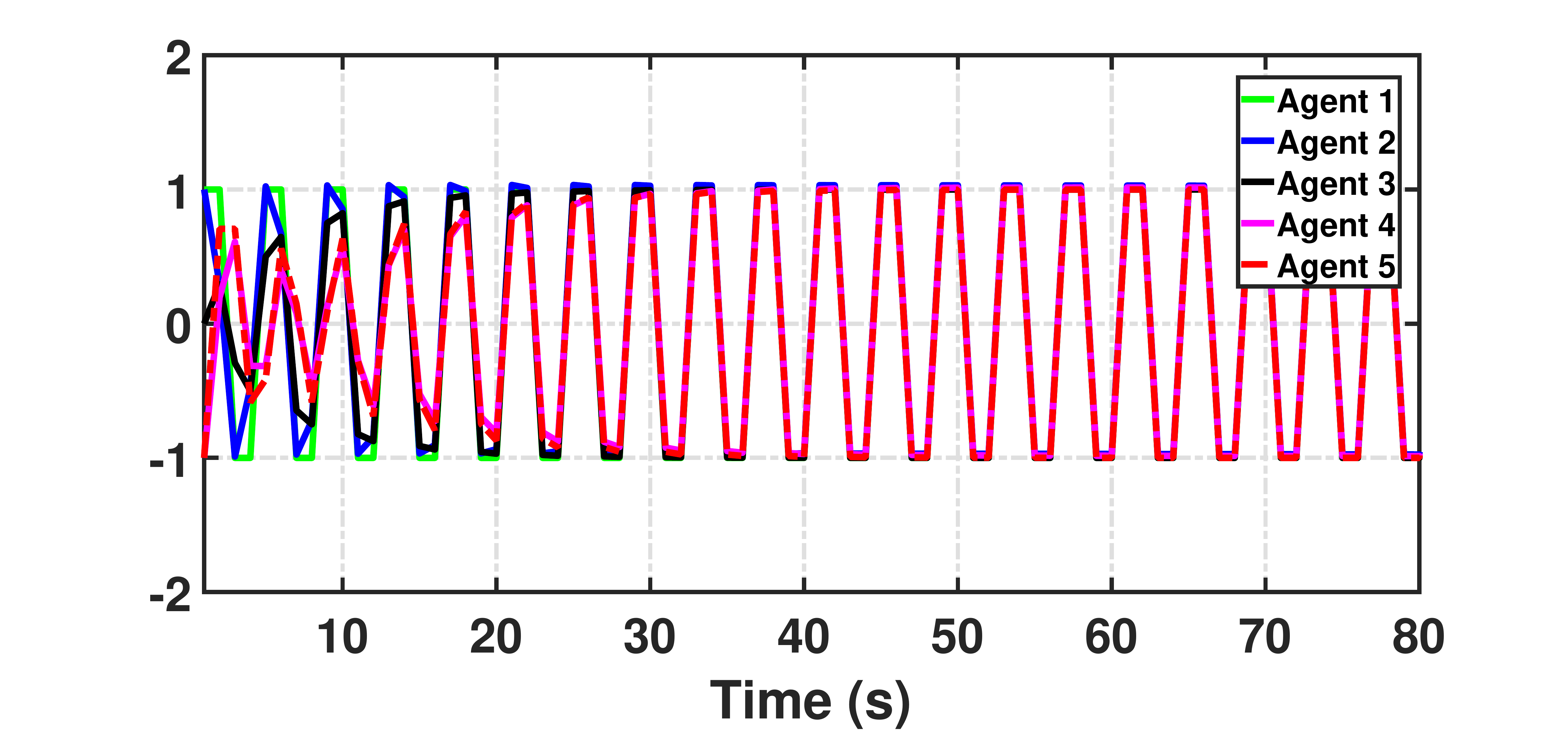}
\caption{}
\end{subfigure} 
\caption{The DMAS response under the effect of IMP-based attack on agent 2 (root node) with adaptive compensator. (a) The agent's state without adaptive compensator  (b) The agent's state with adaptive compensator}
\label{s3}
\vspace{-0.3cm}
\end{figure}

\begin{figure}[H]
\begin{subfigure}[b]{\columnwidth}
\includegraphics[width=\columnwidth,height=45mm]{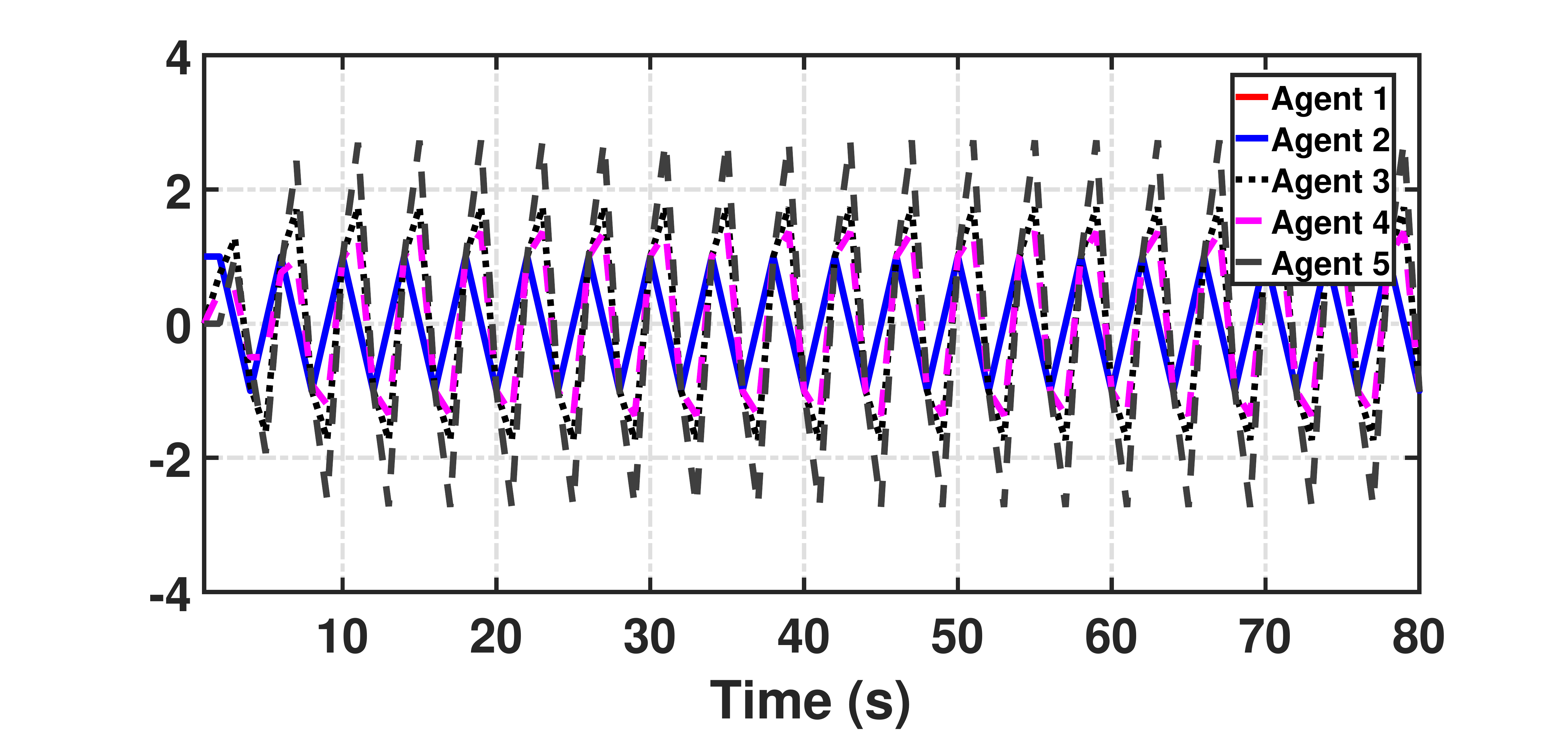} 
\caption{}
\end{subfigure}
\begin{subfigure}[b]{\columnwidth}
\includegraphics[width=\columnwidth,height=45mm]{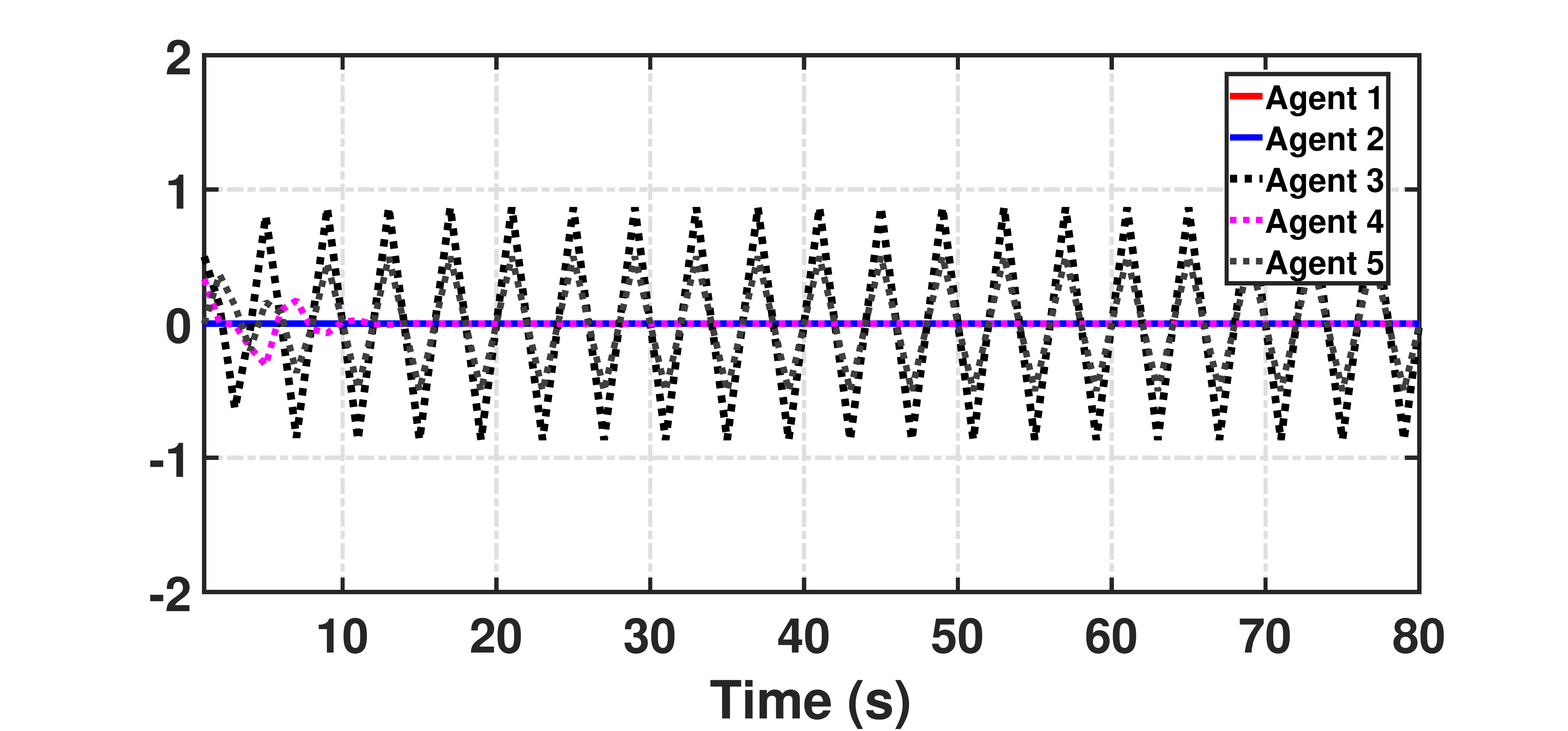}
\caption{}
\end{subfigure} 
\caption{The DMAS response under the effect of IMP-based attack on agent 3 (non-root node) without adaptive compensator. (a) The agent's state  (b) The local neighborhood tracking error}
\label{s1}
\end{figure}

\begin{figure}[H]
\begin{subfigure}[b]{\columnwidth}
\includegraphics[width=\columnwidth,height=45mm]{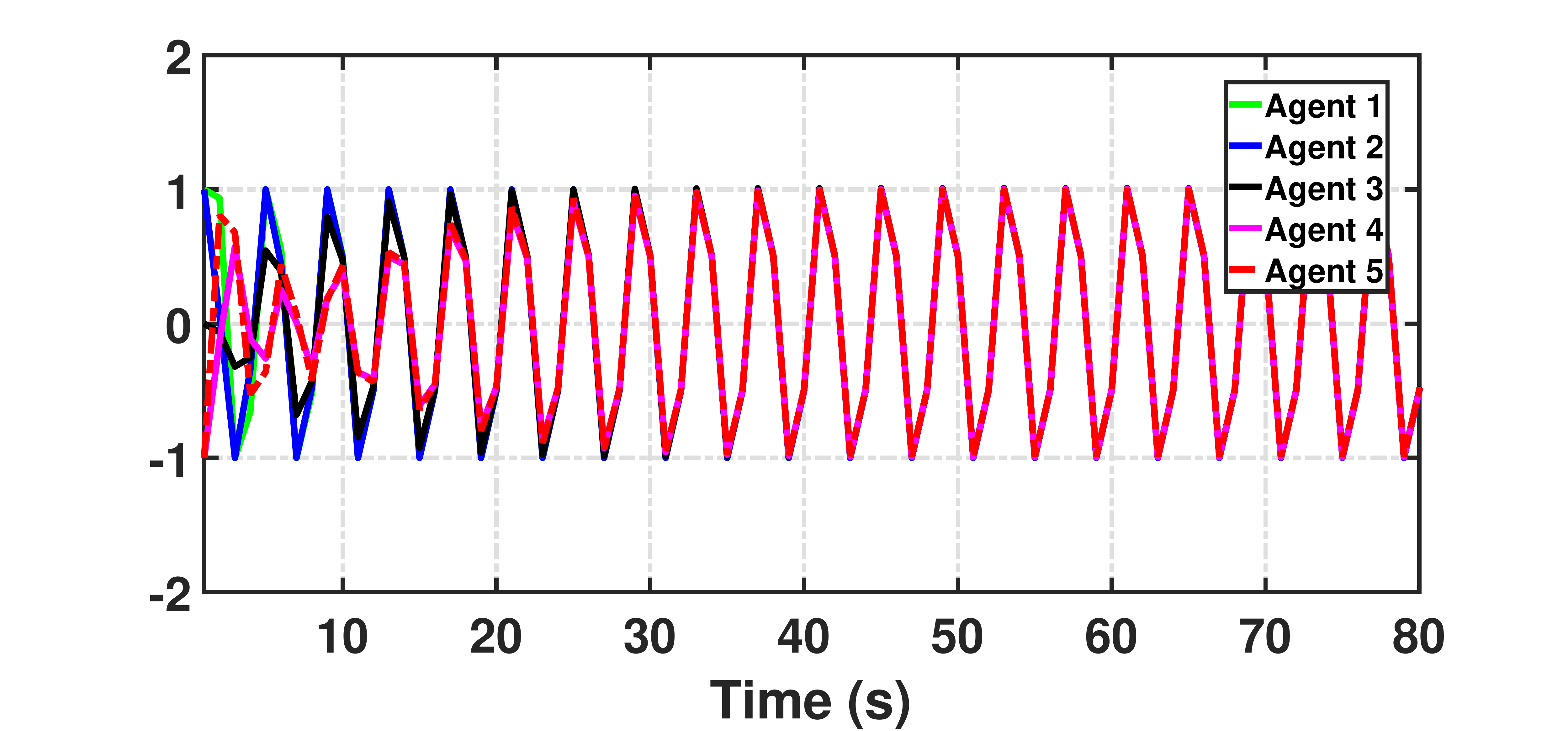}
\caption{}
\end{subfigure}
\begin{subfigure}[b]{\columnwidth}
\includegraphics[width=\columnwidth,height=45mm]{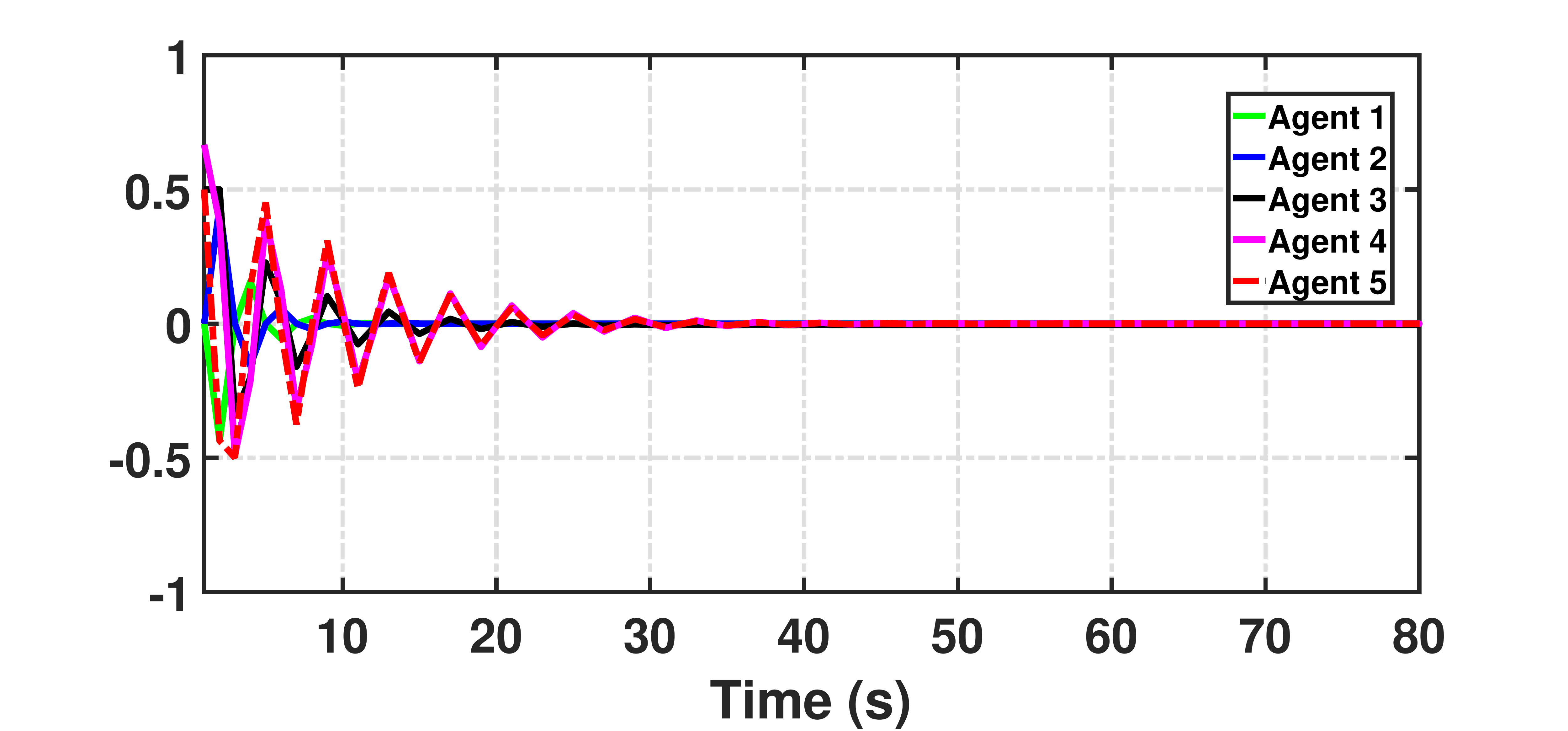}
\caption{}
\end{subfigure}
\caption{The DMAS response under the effect of IMP-based attack on agent 3 (non-root node) with adaptive compensator. (a) The agent's state  (b) The local neighborhood tracking error}
\label{s2}
\end{figure}


\noindent
Consider the effect of attack on actuator of Agent $2$ by IMP-based attack signal i.e. $u_{2}^{a}(k) =sin(k)$.  Fig. \ref{s3}(a) shows that the compromised agent destabilizes the entire network. All agents of the DMAS deviate from the desired consensus behavior. The simulation results verify  Theorem $2$ and  Theorem  $3$. Let $Q_1$ and $R_1$  be identity matrix in (\ref{d7}) and (\ref{d8}), respectively.  Now, the proposed resilient control protocol in  (\ref{d9}) with adaptive attack compensator in  (\ref{d10}) is incorporated and Fig. \ref{s3}(b) shows the response of the system. The system states achieve the desired consensus behavior, even in the presence of the attack on root node. This illustrates the mitigation of sophisticated attack using the designed resilient controller.

Now, we present the results for the effect of the attack on a non-root node.
Consider an IMP-based attack signal is launched on actuator of Agent $3$ (non-root node) i.e. $u_{3}^{a}(k) =sin(k)$.  Fig. \ref{s1}(a) and (b) show that Agents  $4$ and  $5$ which are reachable from the compromised Agent $3$ do not converge to the desired consensus value and the local neighborhood tracking error goes to zero for intact agents. These results comply with Theorem $3$ and Theorem $4$. Then, the resilient control protocol in  (\ref{d9}) with adaptive attack compensator in  (\ref{d10}) is used. Fig. \ref{s2}(a) and (b)  illustrate that the system states achieve the desired consensus behavior and the local neighborhood tracking error goes to zero, even in the presence of the attack  on non-root node $3$. This demonstrates the mitigation of attack using the developed resilient controller.

\vspace{0.0cm}
\section{Conclusion}
This paper presents a rigorous analysis of the effects of attacks for leaderless discrete-time DMAS and designs a resilient distributed control protocol their mitigation. It is shown that the attack on a compromised agent can propagate through the entire network and affects intact agents those are reachable from it. Then, the IMP for the attacker shows that an attack on a single root node can destabilize the entire network. The attacker does not require to know about the communication graph and the system dynamics.  Furthermore, the ineffectiveness of existing robust approach is discussed for sophisticated attacks. To overcome the effect of the attacks on sensor and actuators of the agent in discrete-time DMAS, a  resilient controller is developed based on an expected state predictor.  The presented controller shows that the attack on sensor and actuator can be mitigated without compromising the connectivity of the network and achieves the desired consensus. Although we have considered a general leaderless consensus problem for the proposed controller, it can be used for the other DMAS problems such as leader-follower and containment control problem. The analysis and effectiveness of the presented work have shown in simulation results.

\end{document}